\RequirePackage{fix-cm}
\documentclass[smallextended]{svjour3}       
\smartqed  
\usepackage{graphicx}
\usepackage{amsmath, amssymb, amsfonts}
\usepackage[sort&compress,numbers]{natbib}
\usepackage{booktabs}
\usepackage{hyperref}
\hypersetup{
    colorlinks=true,
    linkcolor=magenta,
    filecolor=blue,      
    urlcolor=blue,
    citecolor=blue,
    pdftitle={Sharelatex Example},
    bookmarks=true,
    }
\usepackage{multirow}
\newtheorem{mydef}{Definition}
\usepackage{tcolorbox}
\begin{document}

\title{A Survey on Mining and Analysis of Uncertain Graphs
\thanks{The author is supported by the Start Up Research Grant provided by Indian Institute of Technology Jammu, India.}
}


\author{Suman Banerjee         
}


\institute{Suman Banerjee \at
              Department of Computer Science and Engineering \\
              Indian Institute of Technology Jammu, India.\\              
              \email{suman.banerjee@iitjammu.ac.in}           
}

\date{Received: date / Accepted: date}

\maketitle

\begin{abstract}
\emph{Uncertain Graph} (also known as \emph{Probabilistic Graph}) is a generic model to represent many real\mbox{-}world networks from social to biological. In recent times analysis and mining of uncertain graphs have drawn significant attention from the researchers of the data management community. Several noble problems have been introduced and efficient methodologies have been developed to solve those problems. Hence, there is a need to summarize the existing results on this topic in a self\mbox{-}organized way. In this paper, we present a comprehensive survey on uncertain graph mining focusing on mainly three aspects: (i) different problems studied, (ii) computational challenges for solving those problems, and (iii)  proposed methodologies. Finally, we list out important future research directions.
\end{abstract}

\section{Introduction} \label{intro}
Graphs are often used to represent real\mbox{-}world networks, such as \emph{social networks} (nodes represent users and edges represent social ties) Carrington et al. \cite{carrington2005models}, \emph{transportation networks} (nodes represent cities and edges represent roads) Bell and Iida \cite{bell1997transportation}, \emph{protein-protein interaction networks} (nodes represent proteins and edges represent interaction relationship) Brohee and Van Helden \cite{brohee2006evaluation} and so on. In many realistic applications, \emph{uncertainty} is intrinsic in graph data due to many practical reasons, which includes noisy measurement Aggarwal \cite{aggarwal2010managing}, inferences and prediction error Adar and Re \cite{adar2007managing}, explicit manipulation and so on. As an example, in case of protein\mbox{-}protein interaction (PPI) networks, as high throughput interaction detection methods are often erroneous, and hence interaction of two proteins is probabilistic in nature. This kind of graphs are often called as uncertain graphs or probabilistic graphs, where each edge of the network is associated with a probability value. The interpretation of this probability may vary from context to context. In case of PPI Network, the assigned probability to an edge is basically its existential probability. On the other hand, in case of a social network,  the assigned probability to an edge may signify the probability with which one can influence other Chen et al.  \cite{chen2013information}. In a generic seance, an uncertain graph is basically a graph whose edges are marked with a probability value. Due to the wider applicability of uncertain graphs in different domains such as \emph{social network analysis} Kempe et al. \cite{kempe2003maximizing}, \emph{computational biology} Zhao et al. \cite{zhao2014detecting}, \emph{crowed sourcing} \cite{ke2018demonstration, yalavarthi2017select, yalavarthi2017probabilistic}, \emph{recommender systems} \cite{taranto2012uncertain}, \emph{wireless networks} \cite{coon2018conditional} analysis of mining of such networks become key research area in recent times.
\par Extensive studies on uncertain graphs lead to many interesting problems \cite{khan2018uncertain, DBLP:conf/algocloud/KassianoGPT16}, which includes \emph{frequent pattern matching} \cite{yuan2016efficient, chen2018efficient}, \emph{subhraph extraction} \cite{chen2010continuous, yuan2011efficient}, \emph{clustering of uncertain graphs} \cite{ceccarello2017clustering, han2019efficient}, \emph{motif counting} \cite{ma13linc}, \emph{reliability computation} \cite{khan2018conditional} and many more. Also, over the years several solution methodologies have been developed. So there is a need to organize the existing results in a self contained manner. In this paper, we serve this purpose by surveying the existing literature. First, we report the goals of this survey.
\subsection{Focus and Goal of the Survey}
In this survey, the main focus are three folded and they are listed below:
\begin{itemize}
\item Different problems studies on uncertain graph mining and analysis,
\item Major challenges for solving these problems,
\item Proposed solution methodologies.
\end{itemize}
The goal of the survey are as follows:
\begin{itemize}
\item to provide a comprehensive background on uncertain graph mining and different problems introduced and studied in this domain.
\item to propose a taxonomy for classifying the existing literature and brief it in a concise manner.
\item to summarize the existing literature and points out future research directions.
\end{itemize} 
\subsection{Proposed Taxonomy}
\begin{figure}
\rotatebox[origin=c]{90}  {\includegraphics[scale=1.0]{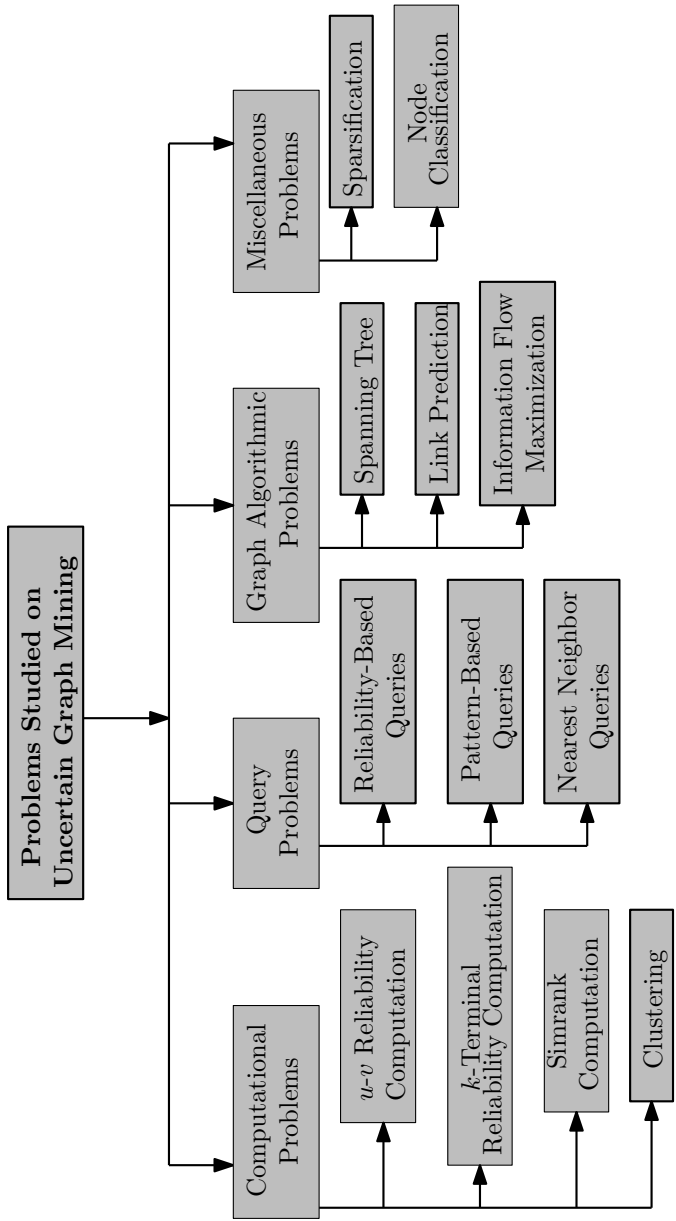}}
  \caption{Taxonomy for classifying the existing literature.}
  \label{fig:Taxonomy}
\end{figure}

Broadly, the problems that have been studied in uncertain graph mining domain can be classified into three main categories: (i) Computational Problems (e.g. computation of $u-v$ reliability, clustering etc.), (ii) Querying Problems (e.g., reachability queries , queries regarding the existence of a particular combinatorial structures etc.), and (iii) Graph Algorithmic Problems (e.g., construction of spanning tree, link prediction, information flow maximization etc.). Also, there are some problems such as \emph{sparsification}, \emph{node classification} etc. which does not come under any of the three heads, and we put them under miscellaneous problems. Figure \ref{fig:Taxonomy} gives a diagrammatic view for the proposed classification of the existing literature on uncertain graph mining.

\subsection{Organization of the Survey}
Rest of the paper is organized as follows: Section \ref{Sec:Pril} lists required preliminary definitions. Section \ref{Sec:Problems} describes different problems studied on uncertain graph mining. In Section \ref{Sec:Challenges}, we report the existing challenges for solving the problems. Existing solution methodologies for these problems are described in Section \ref{Sec:Method}. In Section \ref{Sec:Gaps}, after analyzing the existing literature we describe the current research trends and point out existing research gaps. Section \ref{Sec:ReDir} list out existing future research directions. Finally, Section \ref{Sec:Conclusion} concludes the survey.

\section{Preliminaries} \label{Sec:Pril}
In this section, we describe required preliminary concepts. Initially, we start with the uncertain graph.
\begin{mydef}[Uncertain Graph] \footnote{Please don't confuse between uncertain graph and random graph \cite{bollobas2001random}. Uncertain graph and probabilistic graph are same, however, random graph is completely different and noting to do in this paper. Hence, we have not defined it in this paper.} \label{Def:UG}
An uncertain graph is denoted as $G(V, E, P)$, where $V(G)=\{v_1, v_2, \ldots, v_n\}$ denotes the set of vertices, $E(G)=\{e_1, e_2, \ldots, e_m\}$ denotes the set of edges, and $P$ is the edge weight function that assigns each edge to its probability, i.e., $P:E(G) \longrightarrow (0,1]$.
\end{mydef}
If the uncertain graph is weighted, along with $V$, $E$, and $P$, there is also an edge weight function $W$ that assigns each edge to a real number, i.e., $W:E(G) \longrightarrow \mathbb{R}$. We denote the number of vertices and edges of $G$ as $n$ and $m$, respectively. Depending upon the situation, an uncertain graph may be directed or undirected. As in most of the existing studies, the considered uncertain graph is undirected, in this paper also unless otherwise stated by uncertain graph we mean it is undirected. Also, in some situations instead of single probability value, there may be multiple values associated with an edge. Take the example of a social network, where probability associated with each edge is basically the influence probability between two users. Now, influence probability may vary from context to context \cite{chen2016real}. This means a sportsman can influence his friends and followers regarding any news related to sports with higher probability compared to the others. Let, $C=\{c_1, c_2, \ldots, c_k\}$ be the set of $k$ different contexts. In this case the social network can be modeled as an uncertain graph, where each edge of the network is associated with $k$ number of probability values. In that case, the probability function can be defined as $P: E \times C \longrightarrow (0,1]$. Standered graph theoretic terminologies such \emph{degree}, \emph{neighborhood}, \emph{path}, \emph{subgraph}, \emph{subgraph isomorphism}, \emph{spanning tree} etc. with their definitions and notations have been adopted from  \cite{diestel2012graph} and not described here. For any arbitrary edge $e \in E(G)$, $P(e)$ denotes the probability associated with the edge $e$. \emph{Possible World Semantic} is widely used to represent to an uncertain graph as a probability distribution over a set of deterministic graphs, which is defined next.
\begin{mydef}[Possible World Semantic] \label{Def:Possible}
By this model, an uncertain graph is represented as a probability distribution over $2^{m}$ number of deterministic graphs by keeping an edge with probability $P(e)$ and removing it with probability $(1-P(e))$. Hence, given an uncertain graph $G(V,E, P)$, the probability that the deterministic graph $\mathcal{G}(V,\mathcal{E})$ (Here, $V(G)=V(\mathcal{G})$, and $E(\mathcal{G}) \subseteq E(G)$) will be generated can be computed by the following equation:
\begin{equation}
P_{\mathcal{G}\sqsubseteq G}= \underset{e \in E(\mathcal{G})}{\prod} P(e)  \underset{e \in E(G) \setminus E(\mathcal{G})}{\prod} (1-P(e))
\end{equation}
\end{mydef}

\begin{figure}
\rotatebox[origin=c]{90}  {\includegraphics[scale=1.0]{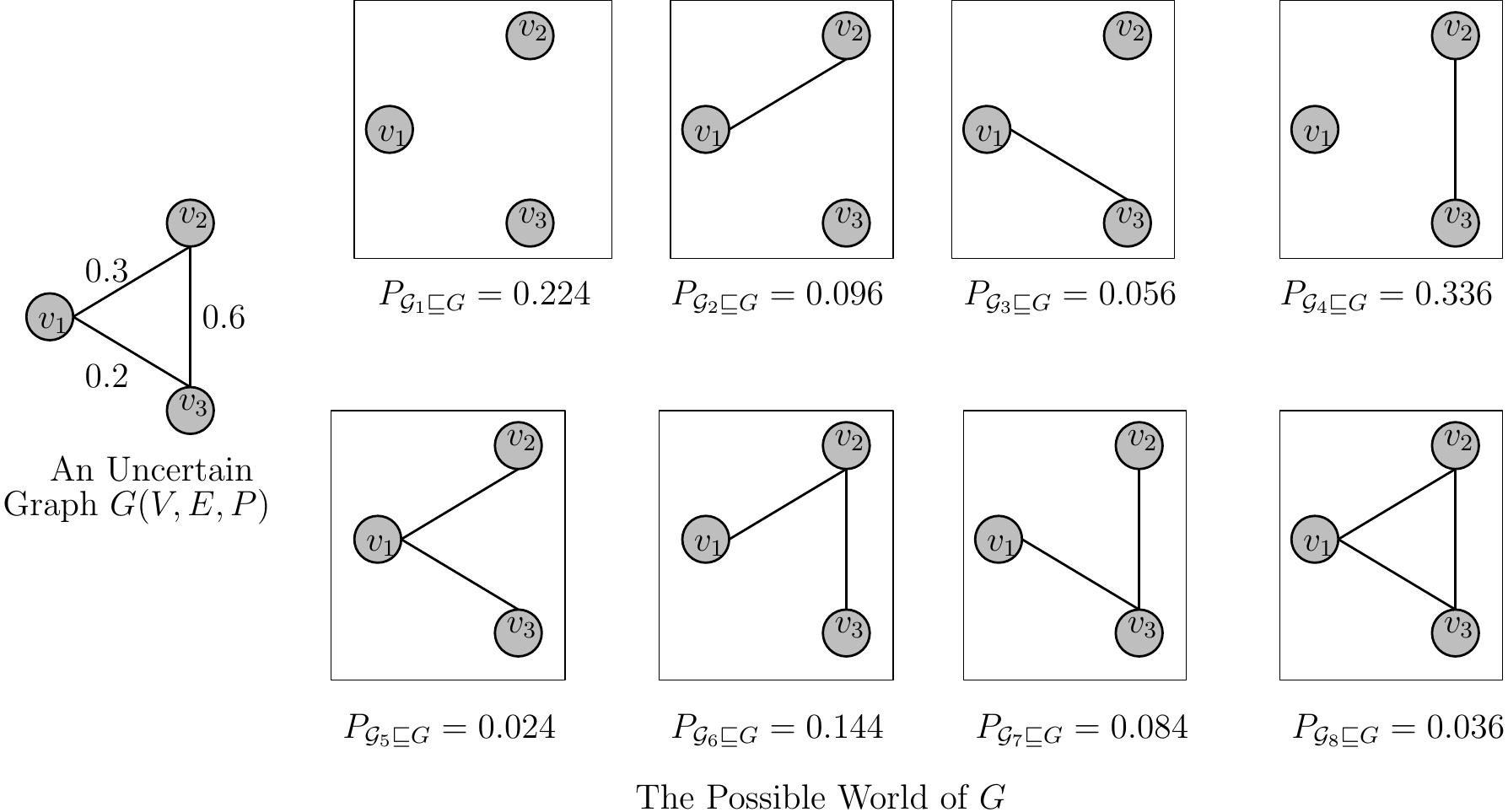}}
  \caption{A toy example of an uncertain graph with its possible world}
  \label{fig:Possible_World}
\end{figure}

Figure \ref{fig:Possible_World} shows an example of an uncertain graphs with its possible world. In the example $|E(G)|=3$, hence $2^{3}=8$ deterministic graphs will be in the possible world of $G$.

Given a graph and two of its vertices distance between them is defined as the length of the shortest path. However, in case of uncertain graphs distance between two vertices can be defined in many ways, which is stated next.
\begin{mydef}[Distance in Uncertain Graphs] \cite{potamias2009nearest}
Given an uncertain graph $G(V, E, P)$, and two of its vertices $u,v \in V(G)$, there are four different distance measure available in the literature:
\begin{itemize}
\item \textbf{Majority Distance}: It is defined as the most probable shortest path distance between $u$ and $v$.
\item \textbf{Expected Distance}: It is defined as the expected shortest path distance between $u$ and $v$ among all the possible world graphs.
\item \textbf{Expected Reliable Distance}: It is defined as the expected shortest path distance between $u$ and $v$ among all the possible world graphs in which $u$ and $v$ are connected.
\item \textbf{Median Distance}: It is defined as the median shortest path distance among all the possible world graphs.
\end{itemize}
Mathematical expressions for computing these distance measures are given in Table \ref{tbl:diatance_summary}. Here, $P_{uv}(d)$ denotes the probability that the distance between $u$ and $v$ is $d$.
\begin{table}[!h]
\centering
 \begin{tabular}{|c|c|}\hline
 Distance Metric & Mathematical Expression \\\hline\hline
 Majority Distance & $dist_{MD}=\underset{d}{argmax}P_{uv}(d)$ \\\hline
 Expected Distance & $dist_{ED}=\underset{d}{\sum}d.P_{uv}(d)$ \\\hline
 Expected Reliable Distance & $dist_{ERD}= \underset{d| d < \infty}{\sum} d \frac{P_{uv}(d)}{1-P_{uv}(d)}$\\\hline
 Median Distance & $dist_{MD}= \underset{D}{argmax} \{\sum_{d=0}^{D}P_{uv}(d)< \frac{1}{2}\}$ \\\hline
 \end{tabular}
 \caption{Different distance measures in uncertain graphs.}\label{tbl:diatance_summary}
\end{table}
\end{mydef}
In a deterministic graph, two vertices are always reachable from each other if they belong to the same connected component. However, in case of uncertain graphs, reachability between two vertices are always probabilistic, which is called as \emph{reliability} and it is defined next. 

\begin{mydef}[$u-v$ Reliability]
Given an uncertain graph $G(V, E, P)$, and two vertices $u,v \in V(G)$, let $I_{\mathcal{G}}(u,v)$ be the indicator boolean variable which takes the value $1$ if $u$ and $v$ are reachable from each other in the deterministic graph $\mathcal{G}$, otherwise it is $0$. Now, reliability between $u$ and $v$ is basically the expected value of rechability, which can be given by the Equation \ref{Eq:2}

\begin{equation} \label{Eq:2}
R(u,v)= \underset{\mathcal{G} \sqsubseteq G}{\sum} I_{\mathcal{G}}(u,v). P_{\mathcal{G} \sqsubseteq G}
\end{equation} 
\end{mydef}

Now, the notion of $u-v$ reliability for a uncertain graph can be extended for a subset of the vertices (called as terminal vertices) which is stated in Definition \ref{Def:NR}. 

\begin{mydef}[Network Reliability] \label{Def:NR}
 For a given  uncertain graph $G(V, E, P)$, and a set of terminal vertices $\mathcal{T}\subseteq V(G)$, the network reliability is defined as the probability that the graph induced by the vertex set $\mathcal{T}$ is connected in $G$. This can be computed by Equation \ref{Eq:6}. Here, $I(\mathcal{G}, \mathcal{T})$ is a indicator variable whose value is $1$ when $\mathcal{T}$ is connected and $0$ otherwise.

\begin{equation} \label{Eq:6}
R(G,\mathcal{T})= \underset{\mathcal{G} \in L(G)}{\sum}I(\mathcal{G},\mathcal{T}). P_{\mathcal{G} \sqsubseteq G}
\end{equation}
\end{mydef}

Subsequently, the notion of $u-v$ reliability has been generalized by Khan et al. \cite{khan2018conditional} when the uncertain graph has more than one probability value in every edge and this is defined next.

\begin{mydef}[Conditional $u-v$ Reliability] \label{Def:CR} \cite{khan2018conditional}
Let, $G(V, E, P)$ be an uncertain graph, and $C=\{c_1, c_2, \ldots, c_k\}$ be the $k$ different contexts. Here, $P$ is defined as $P:E \times C \longrightarrow (0,1]$. Now, for any given $C_{1} \subseteq C$ and with the assumption that the contexts are independently aggregated, the effective probability for any edge $e \in E(G)$ can be computed by the Equation \ref{Eq:3}.
\begin{equation} \label{Eq:3}
P(e| C_{1})= 1- \underset{c \in C_{1}}{\prod} (1- P(e|c))
\end{equation}
Similarly, by considering the contexts $C_{1}$, the probability that the deterministic graph $\mathcal{G}$ will be generated can be computed by the Equation \ref{Eq:4}.
\begin{equation} \label{Eq:4}
P(\mathcal{G}|C_{1})=\underset{e \in E(\mathcal{G})}{\prod} P(e|C_{1})  \underset{e \in E(G) \setminus E(\mathcal{G})}{\prod} (1-P(e|C_{1}))
\end{equation}
Now, given two vertices $u,v \in V(G)$, and given the set of contexts $C_{1}$ the conditional reliability between $u$ and $v$ can be defined by Equation \ref{Eq:4}.
\begin{equation}\label{Eq:4}
R((u,v)|C_{1})= \underset{\mathcal{G}\sqsubseteq G|C_{1}}{\sum} I_{\mathcal{G}}(u,v). P(\mathcal{G}|C_{1})
\end{equation}
Here, $I_{\mathcal{G}}(u,v)$ is an indicator variable whose value is $1$ if $u$ and $v$ is connected in $\mathcal{G}$. 
\end{mydef}
Given two deterministic graphs $\mathcal{G}$ and $\mathcal{H}$, a well known computational task is to determine whether $\mathcal{G}$ contains a subgraph isomorphic to $\mathcal{H}$. This is the popular \emph{subgraph isomorphism problem} \cite{lee2012depth} which is known to be NP-Complete \cite{lewis1983computers}. However, this problem can be translated in case of uncertain graphs by incorporating the concept of \emph{support}, which is defined next.
\begin{mydef}[Support]
Given an uncertain graph $G(V,E,P)$, and a subgraph $S$, the support of $S$ is defined as the fraction of the possible worlds that contains $S$ as subgraph. Mathematically, this can be expressed as follows:
\begin{equation}
sup(S;G)= \frac{|\mathcal{G} \in L(G):S \sqsubseteq \mathcal{G}|}{2^{m}}
\end{equation}
\end{mydef}

Now, given a support value $\alpha \in (0,1)$, an uncertain graph $G(V,E,P)$ and a deterministic graph $\mathcal{G}(V, E^{'})$, one immediate question is that: ``Does the uncertain graph $G$ contains the $G$ as a subgraph with support greater than or equal to $\alpha$?" This kind of querying problems are called structural pattern\mbox{-}based queries. Section \ref{Sec:Querying_Problems} discusses it in more details.  
\par For a given graph, computation of \emph{node similarity} (how similar two nodes are?) remains a central question in graph mining. Among many, one of the popularly adopted node similarity measure is \emph{Simrank} initially proposed by Jeh and Widom \cite{jeh2002simrank}. The intuition behind this similarity measure is that two vertices are similar if they are referenced by similar vertices.
\begin{mydef}[Simrank]
Given a deterministic graph $\mathcal{G}(V,E)$, and two vertices $u,v \in V(\mathcal{G})$, the Simrank score between these two vertices are given by the following equation:
\[
    \sigma(u,v)= 
\begin{cases}
    1 ,& \text{if } u = v\\
    \frac{c}{|N^{in}(u)|. |N^{in}(v)|} \underset{w \in N^{in}(u)}{\sum} \underset{w^{'} \in N^{in}(v)}{\sum} \sigma(w, w^{'}),              & \text{if } u \neq v
\end{cases}
\]
where $N^{in}(u)$, and $N^{in}(v)$ denotes the set of incoming neighbors of the nodes $u$ and $v$, respectively. $c \in (0,1)$ is the decaying factor whose value is chosen as 0.8 \cite{li2010fast}. Section \ref{SubSec:Simrank} summarizes the existing literature for simrank computation on uncertain graphs.
\end{mydef} 

\begin{mydef}[Random Walk on Graphs] \cite{gobel1974random}
This is a discrete time process defined on graphs and also it is one kind of graph search technique. Suppose, at time $t=0$ an object is placed on the vertex $v$ of an undirected graph $\mathcal{G}(V,E)$. At every discrete time steps $t=1, 2, \ldots$ the object must move from one vertex to one of its neighboring one. So, if the object is on the vertex $v_i$ at time $t$, then the probability that the object will be on the vertex $v_j$ at time $t+1$ is given by the following equation.
\[
    m_{ij}= 
\begin{cases}
    \frac{1}{deg(v_i)} ,& \text{if } (v_iv_j) \in E(\mathcal{G})\\
    0,              & \text{if } (v_iv_j) \notin E(\mathcal{G})
\end{cases}
\]
This probability values are reported as a $n \times n$ matrix $M$.
\end{mydef} 
Existing solution methodologies for the simrank computation on uncertain graphs use random walk concept as described in Section \ref{SubSec:Simrank}. Table \ref{tab:Symbols} describes the notations and symbols with their interpretations used in this paper. Many of the symbols have not been introduced yet and done as and when needed.
\begin{table}[h]
  \caption{Notations used in this study}
  \label{tab:Symbols}
  \begin{tabular}{||c|c||}
    \toprule
   Notation & Meaning \\
    \midrule
    $G(V, E, P)$ & An uncertain graph \\
    \hline
    $G(V, E, P, W)$ & A weighted uncertain graph \\
    \hline
    $V(G), E(G)$ & Vertex set and edge set of $G$\\
    \hline 
    $n,m$ & The size of the vertex set and edge set of $G$\\
    \hline 
    $d$ & Diameter of $G$\\
    \hline 
    $P$ & Edge weight function, i.e., $P:E(G) \longrightarrow (0,1]$\\
    \hline
    $W$ & Edge weight function, $W: E(G) \longrightarrow \mathbb{R}^{+}$\\
    \hline
    $A$ & Adjacency matrix of $G$\\
    \hline
    $u,v$ & Any two arbitrary vertices of $V(G)$\\
    \hline
    $N^{in}(u)$ &  Set of incoming neighbors of $u$\\
    \hline
    $e / (uv)$ & An arbitrary edge from $E(G)$\\
    \hline
    $P(e)/ P(uv)$ & Edge probability of the edge $e=(uv)$\\
    \hline
    $L(G)$ & Possible world set of $G$\\
    \hline
    $P_{\mathcal{G}\sqsubseteq G}$ & Generation probability of $\mathcal{G}$ from $G$\\
    \hline
    $\mathcal{G}(V, \mathcal{E})$ & A deterministic graphs from $L(G)$\\
    \hline
    $\sigma(u,v)$ & Simrank between $u$ and $v$ in $G$\\
    \hline
    $\mathcal{T}$ & Set of terminal vertices from $V(G)$\\
    \hline
    $\eta$ & Reachability / Reliability threshold \\
    \hline
    $C$ & Set of different contexts/ tags \\
    \hline
    $P(e|c)$ & The edge probability of the edge $e$ for the tag $c$ \\
    \hline
    $C_{1}$ & A subset of contexts $C_{1} \subseteq C$\\
    \hline
    $P(e|C_{1})$ & Edge probability of $e$ after the aggregating the tags in $C_{1}$\\
    \hline
    $R((s,t)| C_{1})$ & Reliability between $s$ and $t$ for the tags in $C_{1}$\\
    \hline
    $D_{\beta}(u,v)$ & Distance between $u$ and $v$ in $G$ under distance measure $\beta$\\
    \hline
    $sup(S; G)$ & Support of $S$ on $G$\\
    \hline
    $\alpha$ & Specification Ratio \\
    \hline
    $S$ & The Simrank matrix \\
    \hline
    $q$ & An $s-t$ query\\
    \hline
    $G(q)$ & Uncertain graph corresponding to the $s-t$ query $q$\\
    \hline
    $g$ & Query Subgraph \\
    \hline
    $k$ & The set $\{1,2, \ldots, k \}$ \\
   \bottomrule
\end{tabular}
\end{table}
\section{Problems Studied on Uncertain Graphs} \label{Sec:Problems}
In these section, we describe the problems that have been studied in the domain of uncertain graph mining. We devote one subsection to each categories of sub-problems as shown in Figure \ref{fig:Taxonomy}.
\subsection{Querying Uncertain Graphs} \label{Sec:Querying_Problems}
Different kinds of querying problems have been studied in the context of uncertain graph such as \emph{k\mbox{-}Nearest Neighbor Queries} (kNN Queries), \emph{Reliability Queries}, \emph{Queries for Structural Pattern} and so on. We describe them one by one.

\subsubsection{Reliability\mbox{-}Based Queries}
Analysis of network reliability is old still active research area \cite{ball1986computational, brecht1988lower, ball1995network,  khan2014fast, guo2019polynomial}. Given an uncertain graph $G(V,E,P,W)$, a probability threshold $\eta \in (0,1)$, and a set of source nodes $S \subset V(G)$, \emph{Reliability\mbox{-}Based Query Problem} asks to return $T \subseteq V(G) \setminus S$, such that the probability of reachability from the set $S$ to $u$, $\forall u \in T$ is more than $\eta$ \cite{khan2014fast}. From the computational point of view, the problem can be posed as follows:
\begin{tcolorbox}

\underline{\textsc{Reliability\mbox{-}Based Query Problem}} \\
\textbf{Input:} Uncertain Graph $G(V,E,P,W)$, a vertex subset $ S \subset V(G)$, and probability threshold $\eta$.

\textbf{Question:} Return the set $T \subseteq V(G) \setminus S$, such that $\forall u \in T$, $P(S,u) \geq \eta$.
\end{tcolorbox}

\subsubsection{Conditional Reliability Maximization Queries}
Recently, Khan et al. \cite{khan2018conditional} studied the conditional reliability maximization problem. In case of single source single terminal variant of this problem along with an uncertain graph $G(V, E, P)$ , and set of different contexts $C=\{c_1, c_2, \ldots, c_k\}$ with $P: E(G) \times C \longrightarrow (0,1]$, two specified vertices $s,t \in V(G)$, and a positive integer $\ell \leq k$, this problem asks to choose subset of the contexts $C^{'} \subseteq C$ such that $|C^{'}| =\ell$ such that conditional $u-v$ reliability  $R((s,t)|C^{'})$ as described in Definition \ref{Def:CR} gets maximized. Mathematically, this problem can be expressed by the following equation.

\begin{equation}
C^{*}=\underset{C^{'} \subseteq C; |C^{'}|=\ell}{argmax} R((s,t)|C^{'})
\end{equation}

From the computational perspective, this problem can be posed as follows:

\begin{tcolorbox}

\underline{\textsc{Conditional Reliability Maximization Queries}} \\
\textbf{Input:} Uncertain Graph $G(V,E,P)$, Set of Contexts $C=\{c_1, c_2, \ldots, c_k\}$. with $P: E(G) \times C \longrightarrow (0,1]$, Two Vertices $s,t \in V(G)$, and a Positive Integer $\ell \leq k$.

\textbf{Question:} Return $C^{'} \subseteq C$, such that $|C^{'}|= \ell$ and $R((u,v)|C^{'})$ quantity is maximized.
\end{tcolorbox}

\subsubsection{Nearest Neighbor(NN) Queries} This is a very generic data mining task, where given a set of data points and a specific one of them (say $d$) the question is to return $k$ most similar data points as $d$ from the remaining \cite{roussopoulos1995nearest}. This problem has also been studied extensively in graph data as well \cite{malkov2014approximate}. Recently, the $k$-NN Query Problem has been studied on uncertain graph as well \cite{potamias2010k}. Given a weighted uncertain graph $G(V,E,P,W)$, a specific vertex $v \in V(G)$, a probabilistic distance function $D_{P}$ (anyone of four as shown in Table \ref{tbl:diatance_summary}), and a positive integer $k$, the \emph{kNN Query on Uncertain Graph} Problem asks to return the set $\{u_1, u_2, \ldots, u_k\} \subseteq V(G) \setminus \{v\}$ such that for any vertex in $w \in V(G) \setminus \{v, u_1, u_2, \ldots, u_k\}$, $D_{P}(v,w)\geq D_{P}(v,u_i)$ for any $i \in [k]$. From the computational perspective, the problem can be posed as follows:
\begin{tcolorbox}

\underline{\textsc{$k$-NN Query on Uncertain Graph Problem}} \\
\textbf{Input:} Weighted Uncertain Graph $G(V,E,P,W)$, a specific vertex $v \in V(G)$, a probabilistic distance function $D_{P}$, and $k \in \mathbb{Z}^{+}$.

\textbf{Question:} Return the vertex set $\{u_1, u_2, \ldots, u_k\} \subseteq V(G) \setminus \{v\}$ such that for any vertex in $w \in V(G) \setminus \{v, u_1, u_2, \ldots, u_k\}$, $D_{P}(v,w)\geq D_{P}(v,u_i)$ for any $i \in [k]$.
\end{tcolorbox}
This problem can also be defined on unweighted uncertain graph as well. In that case $\forall e \in E(G)$, $W(e)=1$. Here, we have described the weighted version only. 

\subsubsection{Structural Pattern\mbox{-}Based Queries}
Finding or counting a given structure (also known as query graph) in a graph database is a fundamental graph mining task \cite{yan2005substructure, kuramochi2005finding}. This problem has also been studied in the context of uncertain graph as well. Particularly, the structure that has been studied extensively is the \emph{frequent subgraph} \cite{zou2010discovering, li2012mining}, densest subgraph \cite{zou2013polynomial}, \emph{Clique} \cite{mukherjee2015mining, DBLP:journals/tkde/MukherjeeXT17, zou2010finding} etc.
\begin{tcolorbox}
\underline{\textsc{Reliability\mbox{-}Based Query Problem}} \\
\textbf{Input:} Uncertain Graph $G(V,E,P)$, a structure $X$, and a probability threshold $\alpha$.\\
\textbf{Question:} Is there a pattern $X$ whose support value is greater than or equal to $\alpha$.
\end{tcolorbox}

Here, $X$ may be a $k$\mbox{-}clique, $k$\mbox{-}truss and many more.

\subsection{Computation on Uncertain Graphs}
Different computational problems have been studied in the context of uncertain graphs. Here, we list them one by one.
\subsubsection{$u-v$ Reliability Computation:} Given an uncertain graph $G(V,E,P)$, and two vertices $u$ and $v$, this problem asks to compute the probability that $v$ is reachable from $u$. It can be considered as the fundamental rechability problem in uncertain graph context. From the computational perspective, the problem can be list as follows:
\begin{tcolorbox}

\underline{\textsc{$u-v$ Reliability Computation}} \\
\textbf{Input:} An Uncertain Graph $G(V,E,P)$, two specific vertices $u, v \in V(G)$.

\textbf{Task:} Compute the reliability between $u$ and $v$, i.e., $R(u,v)$ in $G$.
\end{tcolorbox}
A variant of this problem has been introduced by Jin et al.\cite{jin2011distance}, where along with $G(V,E,P)$,  $u, v \in V(G)$, we are also given with a distance $d$ and the question is that what is the probability that $u$ is connected with $v$ within the distance $d$.
\subsubsection{$k$\mbox{-}Terminal Reliability Computation:} This is a more generalized version of the $u\mbox{-}v$ reliability computation problem. In this problem, we are given with a uncertain graph $G(V,E,P)$ and a set of $k$ terminal vertices $\mathcal{T}$, the network reliability $R(G,\mathcal{T})$ can be given by the following equation:
\begin{equation}
R(G,\mathcal{T})= \underset{\mathcal{G} \in L(G)}{\sum}I(\mathcal{G},\mathcal{T}). P_{\mathcal{G} \sqsubseteq G}
\end{equation}
where $I(\mathcal{G},\mathcal{T})$ is a indicator variable whose value will be $1$, if the vertices in $\mathcal{T}$ are connected in $\mathcal{G}$ and $0$ otherwise. This problem assks to compute the quantity $R(G,T)$ \cite{sasaki2019efficient}. From the computational point of view, this problem can be posed as follows:

 \begin{tcolorbox}
\underline{\textsc{Network Reliability Computation:}} \\
\textbf{Input:} An Uncertain Graph $G(V,E,P)$, a subset of vertices $\mathcal{T}$.

\textbf{Task:} Compute the network reliability $R(\mathcal{T},G)$ in $G$.
\end{tcolorbox}

\subsubsection{Simrank Computation:} Simrank is a popular similarity measure between any two vertices of a graph due to its applications in different problems including entity resolution \cite{li2010eif, yin2007object}, similar protein identification \cite{whalen2015sequence} and so on. Recently, Simrank computation problem has also been studied for uncertain graph as well \cite{du2015probabilistic, zhu2016simrank, zhu2017simrank}. For a given uncertain graph $G(V,E,P)$, this problem asks to compute a $n \times n$ similarity matrix $S_{sim}$, where the $(i,j)$\mbox{-}th entry contains the SimRank similarity value. Computationally, the problem looks like the following: 

 \begin{tcolorbox}
\underline{\textsc{Simrank Computation on Uncertain Graphs:}} \\
\textbf{Input:} An Uncertain Graph $G(V,E,P)$, a subset of vertices $\mathcal{T}$.

\textbf{Task:} Compute the network reliability $R(\mathcal{T},G)$ in $G$.
\end{tcolorbox}

\subsubsection{Clustering of Uncertain Graphs}
Clustring is a very generic data mining task. Given a set of data points this problem asks to partition them into a number (may be given or may be dataset dependent) of partitions (may be overlapping) such that data points belongs to the same partitions should be similar in some sence and data points belongs to the different partition should be different \cite{jain1999data}. Clustring has also been stuided extensively on graph data \cite{aggarwal2010survey}. This problem has also been  studied in the context of uncertain graph as well \cite{liu2012reliable, han2019efficient}. The problem can be summarized as follows:

\begin{tcolorbox}

\underline{\textsc{Clustering of Uncertain Graph}} \\
\textbf{Input:} An Uncertain Graph $G(V,E,P)$, and the number of clusters $K$.

\textbf{Task:} Partition $V(G)$ into $C_1$, $C_2$, $\ldots$, $C_K$ based on certain similarity measure.
\end{tcolorbox}

\subsection{Graph Algorithmic Problems}

\subsubsection{Spanning Tree Problem:} 
Given a weighted graph, computing the minimum cost spanning tree is classic problem in Graph Algorithms \cite{cormen2009introduction}. This problem has been studied when the edge weights are uncertain \cite{megow2017randomization, erlebach2008computing}. Recently, this problem has also been studied in the context of uncertain graphs, where the weight of the edge is fixed, however there is a probability of exsistence associated with every edge \cite{zhang2016minimum}. From the computational perspective, this problem can be posed as follows:

\begin{tcolorbox}
\underline{\textsc{Most Reliable Spanning Tree Problem on Uncertain Graphs}} \\
\textbf{Input:} An Weighted Uncertain Graph $G(V,E,P, W)$.\\
\textbf{Task:} Compute the minimum cost spanning tree having the highest probability.
\end{tcolorbox}

\subsubsection{Link Prediction Problem:} 
Link prediction is a classic graph mining task, where the snapshot of the network in different times, say $t=0, 1, \ldots, T$, i.e., $G_{0}$, $G_{1}$, $\ldots$, $G_{T}$ is given and the task is to predict the network snapshot at time $T+1$, i.e., $G_{T+1}$ \cite{DBLP:conf/pakdd/ZhangZ19}.
This problem has lot of applications in social network analysis \cite{liben2007link}, recommerder systems \cite{li2009recommendation} and so on. Recently, this problem has has been studied considering the edge uncertainty \cite{ahmed2016efficient, martinez2017survey}. From the the computational point of view, this problem can be posed as follows:

\begin{tcolorbox}
\underline{\textsc{Link Prediction Problem on Uncertain Graphs}} \\
\textbf{Input:}  Snapshots of an Uncertain Graph $G_{t}(V,E_{t},P_{t})$, $\forall t \in [T]$.\\
\textbf{Task:} Output a $n \times n$ similarity matrix, where its $(i,j)$\mbox{-}th denotes the likelihood of $(v_iv_j) \in E(G_{T+1})$.
\end{tcolorbox}

\subsubsection{Information Flow Maximization Problem:}
Network flow is a classic problem in network algorithms, where given a directed graph with each edge is marked with its capasity and a source and target vertex, the goal here is to decide how much flow need to pass through each of the links such that total flow from the source to the target vertex is maximized \cite{ahuja1988network}. Recently, this problem has been studied in the uncertain graph as well, however in a different setting \cite{frey2017efficient, DBLP:conf/icde/FreyZER18}. Here the problem is defined as follows: Given a vertex weighted uncertain graph $G(V,E,P)$, a query vertex $Q \in V(G)$, and a positive integer $k$, this problem asks to find out the subgraph $\mathcal{G}$ that can contain at most $k$ edges that maximizes the expected information flow towards $Q$. From the computational point of view, this problem can be posed as follows: 
\begin{tcolorbox}
\underline{\textsc{Information Flow Maximization Problem on Uncertain Graph}} \\
\textbf{Input:} A Vertex Weighted Uncertain Graph $G(V,E,P,W)$, a query vertex $Q \in V(G)$, a positive integer $k$.\\
\textbf{Task:} Output a $n \times n$ similarity matrix, where its $(i,j)$\mbox{-}th denotes the likelihood of $(v_iv_j) \in E(G_{T+1})$.
\end{tcolorbox}
\subsection{Other Problems}

\subsubsection{Uncertain Graph Sparsification}  \label{SubSubSec:Sparcification}
Due to gigantic size of real\mbox{-}world networks, storing the entire network requires huge storage cost and more importantly querying the entire graph requires huge computational time. The remedy of this problem is not  to store the entire graph. One approach is to resolve this issue is that instead of storing all the edges, store a subset of them. This leads to the problem of graph sparsification. Here, the problem is to decide which edges to store such that some specific property of the graph will not change much in the sparcified graph \cite{fung2019general, spielman2011graph}. Several graph sparcification techniques have been proposed in the literature, such as sparner\mbox{-}based sparscifier \cite{peleg1989graph}, cut\mbox{-}based sparcifier \cite{fung2019general} and many more. Recently, cut\mbox{-}based uncertain graph sparsification problem has been studied by Parchas et al.  \cite{parchas2018uncertain}. Now, this problem has been formalized as follows: Given an uncertain graph $G(V,E,P)$, and a vertex subset $S \subseteq V(G)$, the expected cut size is defined by Equation \ref{Eq:7}.

\begin{equation} \label{Eq:7}
C_{G}(S)= \underset{\substack{e=(uv) \in E(G) \\ u \in S, v \in V(G) \setminus S}} {\sum} P(e)
\end{equation} 

Now, the \emph{Absolute Discrepancy} ($\delta_{A}(S)$) and \emph{Relative Discrepancy} ($\delta_{R}(S)$) for the uncertain graph $G(V,E,P)$ with respect to the sparcified uncertain graph $G^{'}(V,E^{'},P^{'})$ is defined in Equation \ref{Eq:8} and \ref{Eq:9}, respectively.

\begin{equation} \label{Eq:8}
\delta_{A}(S)= C_{G}(S) - C_{G^{'}}(S)
\end{equation}

\begin{equation} \label{Eq:9}
\delta_{A}(R)= \frac{C_{G}(S) - C_{G^{'}}(S)}{C_{G}(S)}
\end{equation}

For a positive integer $k$, the discrepancy of the sparcified uncertain graph $G^{'}$ is defined as the sum of the absolute values discrepancy values for all $k$ sized subsets.
\begin{equation}
\Delta_{k}(G^{'})=\underset{S \subseteq V(G), |S|=k}{\sum} |\delta_{A}(S)|
\end{equation}

Now, for a given uncertain graph $G(V,E,P)$, a positive integer $k$, and a sparsification ratio $\alpha \in (0,1)$, the uncertain graph sparsification problem asks to to construct another uncertain graph $G^{'}(V,E^{'},P^{'})$ such that $|E(G^{'})|= \alpha |E(G)|$ that minimizes the sum of the discrepancies $\underset{i \in [k]}{\sum} \Delta_{i}(G^{'})$. From the computational point of view, this problem can be posed as follows:

\begin{tcolorbox}
\underline{\textsc{Uncertain Graph Sparsification Problem}} \\
\textbf{Input:} An Uncertain Graph $G(V,E,P)$, a positive integer $k$, a sparsification ratio $\alpha \in (0,1)$.\\
\textbf{Task:} Construct another uncertain graph $G^{'}(V,E^{'},P^{'})$ such that $|E(G^{'})|= \alpha |E(G)|$ to minimize $\underset{i \in [k]}{\sum} \Delta_{i}(G^{'})$.
\end{tcolorbox}

\subsubsection{Node Classification in Uncertain Graphs} The problem of classifying the nodes of a network has been studied extensively \cite{bhagat2011node}. Recently, this problem has also been studied in the context of uncertain graphs as well \cite{dallachiesa2014node, kong2013discriminative, han2015uncertain}. This problem can be posed as follows: Given an uncertain graph $G(V,E,P)$, a set of labels $T_{0}$ and subset of its vertices $S \subset V(G)$ are labeled with a labeling function $L:S \longrightarrow T_{0}$, predict the labels of the nodes in $V(G) \setminus S$. Computationally, this problem can be posed as follows:
 
\begin{tcolorbox}
\underline{\textsc{Node Classification in Uncertain Graphs}} \\
\textbf{Input:} An uncertain graph $G(V,E,P)$, label set $T_{0}$, a subset of vertices $S \subset V(G)$ with labeling function $L:S \longrightarrow T_{0}$.\\
\textbf{Task:} Predict the labels of the nodes in $V(G) \setminus S$.
\end{tcolorbox}

There are several other problems that have been studied such as uncertain graph visualization \cite{schulz2016probabilistic, sharara2011g} etc. As the intended audience of this paper is the researchers of the data mining and data management community, hence we are not focusing on these problems. Also, it is important to note that the problems that have been studied on uncertain graphs may be classified in other way also. However, in our classification, the goal was to put the problems that are of similar kind under the same umbrella. Next, we proceed to describe the major challenges for solving these problems.

\section{Challenges for Solving these Problems} \label{Sec:Challenges}
As mentioned in the literature, there are mainly two major challenges described in the following two subsections.
\subsection{Exponential Number of Possible Worlds}
As mentioned in Section \ref{Sec:Pril} that if for a given uncertain graph  with $m$ edges will have $2^{m}$ number of possible worlds. Even for small values of $m$ (say, $m=100$), the number of possible worlds is excessively large ( more than the number of atoms in this universe). Almost all the problems that have been described in Section \ref{Sec:Problems} requires to enumerate all the possible worlds to output the answer accurately. However, due to bounded computational resources it is not possible to consider all the possible worlds. Now, here the challenge is that probabilistically how many samples to consider for the computation to output the result such that the error is bounded by at most  $\epsilon$. Consider the $u-v$ Reliability Computation Problem as described in Section \ref{Sec:Problems}. Let, $R(u,v)$ and $\hat{R}(u,v)$ be the $u\mbox{-}v$ reliability values when all the sample graphs are considered and when all the graphs are not considered, respectively. So the question here is that for a given $\epsilon$ and $\delta$ how many sample graphs to consider such that the following inequity holds: $Pr[|R(u,v) - \hat{R}(u,v)| \leq \epsilon] \geq 1 - \delta$. To address this issue, several effective sampling techniques have been developed such as recursive sampling \cite{jin2011distance}, recursive stratified sampling \cite{li2014efficient, li2015recursive}, lazy propagation sampling \cite{li2015recursive} and many more. Also, Parchas et al.  \cite{parchas2014pursuit, parchas2015uncertain} proposed to generate deterministic representative instances such that underlying graph properties are preserved.

\subsection{Gigantic Size of Real\mbox{-}World Networks} As described, large size of possible world can be handled by approximately answering the result with high probability. However, if the size of network is itself very large, then processing the graph becomes even frther difficult. Recently, to address this issue a framework called \emph{simultaneously processing approach} has been developed \cite{zou2017scalable}. Basically, this method samples a number of possible worlds independently at random, efficiently store them with compact data structures (as many of the sampled possible worlds have common substructure) and simultaneously process the query to generate the results. However, literature in this category is very limited.
\par In the next section, we report the existing solution methodologies for the problems reported in Section \ref{Sec:Problems}.
\section{Exsisting Solution Methodologies} \label{Sec:Method}
In this section, we briefly describe the existing solution methodologies of the problems described in Section \ref{Sec:Problems}. First we start with the computational problems on uncertain graphs.
\subsection{Computational Problems}
\subsubsection{Reliability Computation}
As mentioned previously, the reliability computation problem has been studied in different variants such as two vertex reliability computation ($u-v$ reliability),   For a given uncertain graph, trivial way to compute reliability is to enumerate all the possible worlds and process each of them sequentially. However, this approach leads to huge computational burden. Recently, Sasaki et al. \cite{sasaki2019efficient} proposed an efficient technique to compute the  network reliability, which reduces the number of required possible worlds and compute the bounds on the $k$\mbox{-}terminal  reliability by efficiently constructing \emph{binary decision diagram}, which is basically a directed acyclic graph. Binary decision diagram has been previously used to compute the network reliability on deterministic graphs as well \cite{hardy2007k}. Reported experimental results show that there proposed methodology leads to less variance and less error rate. Jin et al. \cite{jin2011distance} studied the distance constrained $u\mbox{-}v$ reliability problem and their main contribution is to use the \emph{Horvitz\mbox{-}Thomson Type Estimator} and \emph{Recursive Sampling Estimator} that efficiently combines a deterministic computational procedure to boost up the estimation accuracy. Reported results show the superior performance of these two estimators. Recently,  Ke et al. \cite{ke2019depth} performed an in\mbox{-}depth bench marking study for the $s-t$ reliability problem with different sampling strategies, i.e., Monte Carlo (MC) Sampling, \emph{B.F.S. with Indexing} \cite{zhu2015top}, \emph{Recursive Sampling} \cite{jin2011distance}, \emph{Recursive Stratified Sampling} \cite{li2014efficient, li2015recursive}, \emph{Lazy Propagation Sampling} \cite{li2015recursive}, \emph{Indexing via Probabilistic Trees} \cite{maniu2017indexing}. Their goal was to make a comparative study of different sampling strategies to understand their estimator variance, memory usage etc. 
\par The most simplest sampling technique is the MC sampling. In this approach $k$ deterministic graphs $\mathcal{G}_{1}$, $\mathcal{G}_{2}$, $\ldots$, $\mathcal{G}_{k}$ are sampled out from the possible world. Now, assume that $I_{\mathcal{G}_{i}}(s,t)$ for all $i \in [k]$ denotes the boolean variable whose value is $1$ if $s$ and $t$ is reachable from each other and $0$ otherwise. By this method, we have the following estimated reliability:
\begin{equation}
R(s,t) \approx \hat{R}(s,t)= \frac{1}{k} \underset{i \in [k]}{\sum} I_{\mathcal{G}_{i}}(s,t)
\end{equation}
Here, the estimator $\hat{R}(s,t)$ is an unbiased estimator of the $s-t$ reliability, i.e., $\mathbb{E}[\hat{R}(s,t)]= R(s,t)$. Now, it has been shown in \cite{potamias2009nearest} the number of Monte Carlo Samples required to reach the $s-t$ reliability $\eta$ is greater than or equals to $\frac{3}{\epsilon^{2}R(s,t)} ln(\frac{2}{\eta})$, i.e., $Pr[|\hat{R}(s,t)- R(s,t)| \geq \epsilon R(s,t)] \leq \eta$. Now, the traversing each deterministic graph for checking required $\mathcal{O}(m+n)$ time. Hence, the time requirement for reliability estimation using this technique is of $\mathcal{O}(k(m+n))$. 
\par Later Zhu et al. \cite{zhu2015top} proposed an offline sampling scheme which is also space efficient. In this sampling scheme, the given uncertain graph entirely stored without the existence probability of the edges. However, each edge is associated with bit vector of length $k$. For any arbitrary edge $e \in E(G)$, in its associated bit vector, the $i$\mbox{-}th entry is $1$ if the edge $e$ is present in $\mathcal{G}_{i}$. Now, it can be observed that the traversing in this compact graph structure is equivalent to traversing each of the sampled graph in parallel. It is easy to follow that the index building scheme requires $\mathcal{O}(km)$ time and space requirement is of $\mathcal{O}(n+km)$.
\par Jin et al. \cite{jin2011distance} proposed the `recursive sampling' approach that improves over the MC sampling due to the following two reasons. The first one is that: for a given $s-t$ query, if some edges are already missing in a possible world, then it may not be too much relevant whether other set of edges are present or not. Secondly, many possible worlds share a significant number of existing or missing edges. The working principal of this approach is as follows. Starting with the vertex $s$, an extendable edge $e$ incident on $s$ is randomly sampled for $k$ times. Now, the generated samples are divided into two groups: one group containing $e$ and the other one is not containing $e$. Assume, in the first group of samples by the edge $e$ the reachable node is $w$ and now more edges are expandable. This process is repeated for both the groups by picking up an randomly expandable edges and subdividing the groups into smaller ones. A very similar approach was developed by Zhu et al.  \cite{zhu2011bmc} which is called as the \emph{Dynamic MC Sampling} technique.

\par Now, assume that $E_{1} \subseteq E(G)$ and $E_{2} \subseteq E(G)$ be the set of present and non-present edges in one group. Let $G(E_{1}, E_{2})$ denotes the set of possible worlds, where the edges present in $E_{1}$ are present, though the edges of $E_{2}$ are not present. The generation probability of the deterministic graphs present in the group $G(E_{1}, E_{2})$ is given by the following equation:
\begin{equation}
Pr_{G(E_{1}, E_{2}) \sqsubseteq G}= \underset{e \in E_{1}}{\prod} P(e)\underset{e^{'} \in E_{2}}{\prod}(1- P(e^{'}))
\end{equation}

Now, the $s-t$ reliability of the group $G(E_1,E_2)$ is basically the probability that $s$ and $t$ are reachable when a deterministic graph $\mathcal{G}$ from $G(E_1,E_2)$ are appearing and the following equation computes this. 
\begin{equation}
R_{G(E_1,E_2)}(s,t)= \underset{\mathcal{G} \in G(E_1,E_2)}{\sum} I_{\mathcal{G}}(s,t) \times \frac{Pr_{\mathcal{G} \sqsubseteq G}}{Pr_{G(E_1,E_2) \sqsubseteq G}}
\end{equation}
Now, it is easy to verify that for any arbitrary edge $e \in E(G) \setminus (E_1 \cup E_2)$ the following holds:
\begin{equation}
R_{G(E_1,E_2)}(s,t)= P(e).R_{G(E_1 \cup \{e\},E_2)}(s,t)+(1-P(e))R_{G(E_1,E_2 \cup \{e\})}(s,t)
\end{equation}
This process continues till the $E_1$ contains an $s-t$ path with $R_{G(E_1,E_2)}(s,t)=1$ or $E_2$ contains an $s-t$ cut with $R_{G(E_1,E_2)}(s,t)=0$
\par Later, Li et al. \cite{li2015recursive} proposed the recursive stratified sampling technique that works based on the divide-and-conquer paradigm. In this method, the entire probability space is divided into $r+1$ non-overlapping subspaces by selecting $r$ edges. Each division we call as `stratum'. Let $X$ be the set of $r$ edges chosen via breadth first search from the source node $s$. Let, $Y$ is a boolean matrix which stores the status value of the selected edges in different stratum. $Y_{ij}$ will be equal to $1$ if the $j$\mbox{-}th ($1 \leq j \leq r$) edge belongs to the $i$\mbox{-}th stratumor not. Now, a deterministic graph $\mathcal{G}$ from the possible world belongs to the $i$\mbox{-}th stratum is given by the following equation:

\begin{equation}
\pi_{i}= \underset{e_j \in X  \wedge Y_{ij}=1}{\prod} P(e_j) \underset{e_j \in X  \wedge Y_{ij}=0}{\prod} (1-P(e_j))
\end{equation}
The sample size of each stratum is fixed as $k_i=\pi_{i}.k$. Reliability is computed in each of the stratum and the expected value is returned as the $s-t$ reliability. It has been shown in  \cite{li2015recursive} that the time requirement for recursive stratified sampling is same as the MC sampling; i.e.; $\mathcal{O}(k(m+n))$.

\par Li et al. \cite{li2017discovering} proposed the `lazy propagation sampling' scheme. The basic working principle of this method is as follows: ``If the existence probability of an edge is very low, then it quite natural that this edge will not be present in many possible worlds, hence, it is not to probe such edges. To describe formally, this method employs a \emph{geometric distribution} \footnote{\url{https://en.wikipedia.org/wiki/Geometric_distribution}} in each edge and probes an edge if it is activated. It has been shown in \cite{li2017discovering} that the variance of the lazy propagation sampling is same as the MC sampling, though it improves the efficiency.

\par Maniu et al. \cite{maniu2017indexing} proposed the `Indexing via Probabilistic trees' methodology for efficiently computing the $s-t$ reliability in uncertain graphs. This method constructs a tree structure called as the `ProbTree' from the given uncertain graph $G$. When a $s-t$ reliability query $q$ comes, an equivalent graph $G(q)$ is created from the `ProbTree' and the MC sampling is done in $G(q)$ itself. If the size of $G(q)$ is smaller than $G$, then the query $q$ can be evaluated very efficiently. They developed three different indexing schemes, namely, \emph{SPQR trees}, \emph{FWD}, and \emph{LIN ProbTrees}. As per their analysis, among these three, the FWD (fixed width tree decomposition) is the best because the both the \emph{build time} and \emph{query time} of this data structure is linear and the query quality is `lossless' for $\omega \leq 2$, where $\omega$ is the width of the tree decomposition \cite{alber2002improved}.  As per the analysis in \cite{maniu2017indexing}, the time complexity of building the FWD ProbTree is linear in number of nodes of the graph. However, the query execution time is of $\mathcal{O}(k(n^{'}+m^{'}))$. Here, $n^{'}$ and $m^{'}$ denotes the number of nodes and edges present in the graph $G(q)$. The space complexity of this sampling scheme is of $\mathcal{O}(m)$. Table \ref{tab:Time_Space_Sampling} shows the time and space complexity of different sampling strategies for the $s-t$ reliability problem. Table \ref{tab:Sampling} summarizes the pros and cons of different sampling techniques.

\begin{table}[h!]
  \begin{center}
    \caption{Advantages and disadvantages of different sampling techniques for reliability computation}
    \label{tab:Time_Space_Sampling}
    \begin{tabular}{|p{3.9 cm}|p {2.9 cm}|p{2.9 cm}|} 
    \hline
      \textbf{Sampling Technique} & \textbf{Time Complexity} & \textbf{Space Complexity}\\
\hline
\emph{MC Sampling} & $\mathcal{O}(k(m+n))$ & $\mathcal{O}(k(m+n))$ \\
\hline
\emph{B.F.S. with Indexing} & $\mathcal{O}(k(m+n))$ & $\mathcal{O}(n+km)$\\
\hline
\emph{Recursive Sampling} & $\mathcal{O}(nd)$ & $ \mathcal{O} (k(m+n))$\\
\hline
\emph{Recursive Stratified Sampling} & $\mathcal{O}(k(m+n))$ &  $\mathcal{O}(rm)$\\
\hline
\emph{Lazy Propagation Sampling} & $\mathcal{O}(k(m+n))$ & $\mathcal{O}(n)$\\
\hline
\emph{Indexing via Probabilistic Trees} & $\mathcal{O}(k(m+n))$ & $\mathcal{O}(m)$ \\
\hline
\end{tabular}
  \end{center}
\end{table}

\begin{table}[h!]
  \begin{center}
    \caption{Advantages and disadvantages of different sampling techniques for reliability computation}
    \label{tab:Sampling}
    \begin{tabular}{|p{1.7 cm}|p {4.6 cm}|p{4.6 cm}|} 
    \hline
      \textbf{Sampling Technique} & \textbf{Advantages} & \textbf{Disadvantages}\\
      \hline
      \emph{Monte Carlo Sampling} & \begin{itemize}
      \item simple to understand
      \item easy to implement
      \item only one sample is required to store in the main memory
\end{itemize}       & \begin{itemize}
      \item inefficient when the network is large
      \item estimator variance is quite high 
\end{itemize}  \\
      \hline
      \emph{B.F.S. with Indexing} \cite{zhu2015top} & 
      \begin{itemize}
      \item more efficient than MC Sampling
      \item memory requirement is more than MC sampling (to store the bit vectors of the edges)
      \end{itemize}
       & \begin{itemize} 
 \item statistically same variance as the MC Sampling 
 \item same accuracy as the MC Sampling   
      \end{itemize}\\
      \hline
 \emph{Recursive Sampling} \cite{jin2011distance} & 
\begin{itemize}
\item more efficient than MC Sampling
\end{itemize} 
 
  & \begin{itemize}
 \item same variance as MC Sampling
\end{itemize}  \\
      \hline
 \emph{Recursive Stratified Sampling} \cite{li2014efficient, li2015recursive} & 

\begin{itemize}
\item estimator variance is slightly reduced than MC Sampling.
\end{itemize} 
 
  & 
 \begin{itemize}
 \item same variance as MC Sampling
 \item running time is also more than MC and B.F.S. with Indexing.
\end{itemize} \\
      \hline
      \emph{Lazy Propagation Sampling} \cite{li2015recursive} & \begin{itemize}
 \item efficiency improves over the MC sampling
\end{itemize} & 

\begin{itemize}
\item same variance as the MC Sampling
\end{itemize}

\\
      \hline
 \emph{Indexing via Probabilistic Trees} \cite{maniu2017indexing}  & \begin{itemize}
 \item variance is less compared to MC Sampling
\end{itemize} & 

\begin{itemize}
\item Index construction is complicated and takes more time. 
\end{itemize}

 \\
      \hline
    \end{tabular}
  \end{center}
\end{table}

\subsubsection{Simrank Computation} \label{SubSec:Simrank}
To the best of our knowledge, Du et al. \cite{du2015probabilistic} first  study the problem of Simrank computation on uncertain graphs. They proposed a dynamic programming approach to compute the probability values of the probabilistic transition matrix, which works in linear time. To improve the efficiency further, they came up with incremental dynamic programming (IDP) approach. Reported results shows that the IDP approach converges much faster than the naive dynamic programming.
\par Subsequently, Zhu et al. \cite{zhu2016simrank} also studied the same problem. They formalized the notion of random walk in uncertain graphs and also based on this notion they show how one can define Simrank on uncertain graphs. It is important to note that in case of uncertain graphs $k$ step transition probability matrix (i.e., $W^{k}$) is not equal to $k$\mbox{-}th power of the one step transition probability matrix ($W^{1}$). They defined the random walk on uncertain graphs by exploiting the all possible worlds. For a given uncertain graph $G(V,E,P)$, let $X_{0}$, $X_{1}, \ldots, X_{n}$ be the random variable associated with a random walk in $G$. Hence, $Pr_{G}(X_{n}=v_{n} | X_{0}=v_{0}, X_{1}=v_{1}, \ldots, X_{n-1}=v_{n-1})$ can be computed as follows:\\

$Pr_{G}(X_{n}=v_{n} | X_{0}=v_{0}, X_{1}=v_{1}, \ldots, X_{n-1}=v_{n-1})$
\\
$=\underset{\mathcal{G} \in L(G)}{\sum}  Pr_{\mathcal{G}}(X_{n}=v_{n} | X_{0}=v_{0}, X_{1}=v_{1}, \ldots, X_{n-1}=v_{n-1}). Pr_{\mathcal{G} \sqsubseteq G}$.\\

Now, applying the Markovian property,\\

 $=\underset{\mathcal{G} \in L(G)}{\sum}  Pr_{\mathcal{G}}(X_{n}=v_{n} | X_{n-1}=v_{n-1}). Pr_{\mathcal{G} \sqsubseteq G}$\\
 
$ = Pr_{G}(X_{n}=v_{n} | X_{n-1}=v_{n-1})$.\\
 
 Now, $k$\mbox{-}step transition probability from the vertex $u$ to $v$ can be computed as follows: \\
 $Pr_{G}(X_{n}=v|X_{n-k}=u)$\\
 $=\underset{\mathcal{G} \in L(G)}{\sum} Pr_{\mathcal{G}}(X_{n}=v|X_{n-k}=u). Pr_{\mathcal{G} \sqsubseteq G}=\underset{\mathcal{G} \in L(G)}{\sum} Pr_{\mathcal{G}}(u \rightarrow_{k}v). Pr_{\mathcal{G} \sqsubseteq G}$.
 
 Let $W^{k}$ be the $k$\mbox{-}step transition matrix of the uncertain graph $G$. $W^{k}$ can be computed by the following equation.
 \begin{equation}
 W^{k}=\underset{\mathcal{G} \in L(G)}{\sum} Pr_{ \mathcal{G} \sqsubseteq G} \mathbb{W}_{\mathcal{G}}^{k}
 \end{equation}
 where $\mathbb{W}_{\mathcal{G}}^{k}$ denotes the $k$\mbox{-}step transition probability matrix of the possible world $\mathcal{G}$. Based on this random walk on uncertain graphs, they came up with four different approaches to compute Simrank, namely, (i) baseline method, (ii) sampling technique, (iii) two stage method, and (iv)speeding up techniques. Experimental results show that the methods have very high scalability.
 
\subsubsection{Clustering of Uncertain Graphs}
To the best of our knowledge, Liu et al. \cite{liu2012reliable} first studied the clustering problem on uncertain graphs. Suppose, given an uncertain graph $G(V,E,P)$, we want to cluster the graph into $k$ clusters $C_1$, $C_2$, $\ldots$, $C_k$. Consider any possible world $\mathcal{G}_{i} \in L(G)$ and it has $L_{i}$ many connected components denoted as $CC^{i,1}$, $CC^{i,2}$, $\ldots$, $CC^{i,L_{i}}$. The vertices of each connected component are also divided into groups based on their cluster levels. As an example, the vertices of the connected component $CC^{i,j}$ are denoted by $CC^{i,j}_{k}$. First, they introduced the notion of \emph{purity} by the cluster level entropy as follows:
\begin{equation}
F_{p}= \underset{\mathcal{G} \in L(G)}{\sum} Pr_{\mathcal{G} \sqsubseteq G} \underset{j=1}{\sum}^{L_{i}} |CC^{i,j}|. H(\underset{x \in [k]}{\bigcup} CC_{x}^{i,j})
\end{equation}
where $H(\underset{x \in [k]}{\bigcup} CC_{k}^{i,j})=- \underset{x \in [k]}{\sum}\frac{|CC_{x}^{i,j}|}{|CC^{i,j}|} \log (\frac{|CC_{x}^{i,j}|}{|CC^{i,j}|})$ is the entropy of cluster levels for fragment $j$ for the $i$\mbox{-}th possible world of $G$. As mentioned in \cite{liu2012reliable}, if purity is the only criteria then it may happen that the clustering process may leads to a single cluster containing maximum number of nodes. To address this issue, Liu et al. \cite{liu2012reliable} considered the notion of \emph{size balance}, which tells that two clusters can not be too much imbalanced in terms of their sizes. Now, to make the clusters size balanced the following function could be maximized.
\begin{equation}
F_{e}=\underset{\mathcal{G} \in L(G)}{\sum} Pr_{\mathcal{G} \sqsubseteq G}.n. H(\underset{x \in [k]}{\bigcup} C_{x})= n. H(\underset{x \in [k]}{\bigcup} C_{k})
\end{equation}
Here, $H(\underset{k}{\bigcup} C_{k})= - \underset{x \in [k]}{\sum} \frac{|C_{x}|}{n} \log (\frac{|C_{x}|}{n})$. Now, Liu et al. \cite{liu2012reliable} formulated the reliable clustering of an uncertain graph as to minimize the following function:
\begin{equation} \label{Eq:Clus}
F=F_{p}- F_{e}
\end{equation}
They developed a novel $k$\mbox{-}means clustering algorithm to optimize the function mentioned in Equation \ref{Eq:Clus}. Experimental results demonstrate the scalability of this method.
\par Later, Ceccarello et al. \cite{DBLP:journals/pvldb/CeccarelloFPPV17} also studied the uncertain graph clustering problem from a different perspective. Here, the goal is to partition the nodes of the uncertain graph into $k$ cluster in such a way that each of the cluster features a node as a cluster center to maximize the minimum or average reliability (minimum connection probability (MCP) and average connection probability (ACP)) from the cluster center to the other nodes of the cluster. They have shown that this problem is $\#$P\mbox{-}Hard and came up with approximation algorithms for both the MCP and ACP variants of this problem. They have shown that their proposed methodologies generate a $k$\mbox{-}clustering which gives $\Omega((P_{opt-min}(k))^{2})$ and $\Omega((\frac{P_{opt-avg}(k)}{\log n})^{3})$ lower bound in approximation guarantee, where $P_{opt-min}(k)$ and $P_{avg-min}(k)$ denotes the maximum and average connection probability of any $k$\mbox{-}clustering. They compared their results with deterministic weighted graph clustering algorithms to show the efficiency of their proposed methodology.

 \par Kollios et al. \cite{kollios2011clustering} extended the edit distance\mbox{-}based graph clustering technique for uncertain graph. Given two deterministic graphs $\mathcal{G}(V, E_{\mathcal{G}})$ and $Q(V, E_{Q})$, their graph edit distance $ED(\mathcal{G},Q)$ is defined by the following equation.
 \begin{equation} \label{Eq:Edit_Distance}
 ED(\mathcal{G},Q)=| E_{\mathcal{G}} \setminus E_{Q}|+|E_{Q} \setminus E_{\mathcal{G}}|
 \end{equation}
 Now, given an uncertain graph $G(V,E,P)$ and and a deterministic graph $Q(V, E_{Q})$, their graph edit distance $ED(G,Q)$ can be given by the following equation:
 \begin{equation}
 ED(G,Q)=\underset{\mathcal{G} \in L(G)}{\sum}Pr_{\mathcal{G} \sqsubseteq G}. ED(\mathcal{G},Q)
 \end{equation}
 Here, $ED(\mathcal{G},Q)$ can be computed using Equation \ref{Eq:Edit_Distance}. They introduced the notion of \emph{Cluster Graph}, which is basically vertex disjoint disconnected cliques, denoted as $C(V,E_{C})$. That means $V$ is partitioned into $k$ disjoint sets $V_1$, $V_2$, $\ldots$, $V_k$, such that $\forall i \in [k]$, $\forall v, v^{'} \in V_{i}, (vv^{'}) \in E_{C}$, and $\forall i,j \in [k], i \neq j$, $v \in V_{i}$ and $v^{'} \in V_{j}$, $(vv^{'}) \notin E_{C}$. Now, given an uncertain graph $G(V,E,P)$, its clustering problem basically asks to find out the cluster graph $C(V,E_{C})$ such that the edit distance $ED(G,C)$ is minimized. They exploited the connection between this problem with that of correlation clustering and showed that the randomized expected $5$\mbox{-}approximation algorithm proposed by Ailon et al. \cite{ailon2008aggregating} for weighted correlation clustering can be used for solving the uncertain graph clustering problem. Experimental evaluations show that this algorithm generates statistically significant clusters of an uncertain graph and also scale well on real\mbox{-}world networks.
 \par Halim et al. \cite{halim2017efficient} proposed a solution methodology for the uncertain graph clustering problem by exploiting the neighborhood information. By this method first the input uncertain graph is converted into a deterministic graph by classification technique used for edge prediction, and finally deterministic graph clustering technique can be used for clustering. They also performed an experimental study with  different classification technique for edge prediction.
 \par Subsequently, Han et al. \cite{han2019efficient} studied the uncertain graph clustering problem with two different goals: divide the uncertain graph into $k$\mbox{-}clusters such that (i) the average reliability from cluster center to other nodes are maximized (similar to the $k$\mbox{-}median problem), and also (ii) the minimum reliability between any node of the cluster to its cluster center is maximized (similar to the $k$\mbox{-}center problem). Both $k$-center and $k$-median problems has been studied extensively by the researchers of theoretical computer scientists \cite{thorup2005quick, mettu2003online}. For the $k$-median problem, they proposed an $(1-\frac{1}{e})$-factor approximation algorithm, and also for the $k$-center problem they proposed an $(1-\epsilon)OPT_{k}^{c}$ factor approximation (with high probability) algorithm where $OPT_{k}^{c}$ is the optimal value for the $k$-center objective function. Their experimental evaluation shows that proposed approaches significantly out performs the methods proposed by Ceccarello et al. \cite{DBLP:journals/pvldb/CeccarelloFPPV17}.
\par It is important to notice that though there are several clustering techniques for uncertain graphs, however, the criteria is different across the methodologies. Like, in Liu et al.'s \cite{liu2012reliable} study the goal is to generate size balanced clusters, where as in Ceccarello et al.'s \cite{DBLP:journals/pvldb/CeccarelloFPPV17} study the goal is to cluster the uncertain graph to maximize the average/minimum reliability within each cluster. Table \ref{tab:Clustering} briefly summarizes the uncertain graph clustering techniques.
\begin{table}[h!]
  \begin{center}
    \caption{Summery of the Uncertain Graph Clustering Techniques.}
    \label{tab:Clustering}
    \begin{tabular}{|p{1.7 cm}|p {4.0 cm}|p{5.2 cm}|} 
    \hline
      \textbf{Reference} & \textbf{Criteria for Clustering} & \textbf{Main Contributions}\\
      \hline
     Liu et al. \cite{liu2012reliable} & To generate size balanced clustering of an uncertain graph. & \begin{itemize}
      \item Came up with a generalized reliability measurement based on purity and size balance.
      \item Proposed a novel k\mbox{-}means algorithm for uncertain graph clustering.
\end{itemize}       \\
      \hline
      Ceccarello et al. \cite{DBLP:journals/pvldb/CeccarelloFPPV17} & To cluster an uncertain graph such that the average/minimum reliability within each cluster is maximized. & \begin{itemize}
      \item Showed that the problem is $\mathcal{NP}$\mbox{-}Hard.
      \item Proposed methodologies that gives approximation guarantee with respect to optimal $k$\mbox{-}Clustering.
      \end{itemize}\\
      \hline
 Kollios et al. \cite{kollios2011clustering} & To minimize the edit distance between cluster graph and input uncertain graph & \begin{itemize}
 \item Formulated the uncertain graph clustering problem as the general case of \emph{Cluster Edit} Problem.
 \item Showed that the randomized algorithm for proposed by \cite{ailon2008aggregating} can be used to solve the uncertain graph clustering problem
 \end{itemize}\\
      \hline
 Halim et al. \cite{halim2017efficient} & 
 \begin{itemize}
 \item No specific criteria mentioned. 
 \end{itemize}

 & 
 \begin{itemize}
 \item Proposed a methodology that converts the uncertain graph to a deterministic one by applying edge classification technique so that existing deterministic graph clustering techniques can be used.
 \end{itemize}
 
 \\
      \hline
   Han et al.  \cite{han2019efficient} & 
   \begin{itemize}
 \item No specific criteria mentioned. 
 \end{itemize}
   
    & 
    \begin{itemize}
   \item Proposed a methodology that converts the uncertain graph to a deterministic one by applying edge classification technique so that existing deterministic graph clustering techniques can be used.
   \end{itemize}
    
    \\
      \hline
    \end{tabular}
  \end{center}
\end{table}
\par Next, we describe solution methodologies of various graph algorithmic problems that has been studied in the context of uncertain graphs.
\subsection{Querying Uncertain Graphs}
\subsubsection{Querying for Subgraph}
Subgraph searching and querying in a graph database in deterministic setting is a very well studied topic in data management community. However, in probabilistic setting the amount of literature is very limited. Here, we briefly summarize the existing literature.
\paragraph{\textbf{Pattern Matching Queries}} To the best of the author knowledge (also as per the authors' claim) Zou et al. \cite{zou2009frequent} were the first to study the subgraph pattern matching queries on uncertain graphs. As the Subgraph Isomorphism Problem  is NP-Hard even in deterministic graphs, the same hardness result follows in the context of uncertain graphs as well. Now, as the rechability problem is $\#$P-Hard in uncertain graphs, hence the subgraph pattern matching in uncertain graph is also $\#$P-Hard. So, the goal here was to find an approximate solution for this problem. They proposed an efficient approach  to check whether a subgraph should be returned as a solution or not. They also derived a sample bound which is $\frac{16.n. \log(\frac{2}{\delta})}{\epsilon^{2}. minsup}$ such that the $Pr(|\hat{p}-Pr(F)| \geq \epsilon) \leq \delta$. Here, $\hat{p}$ is an estimator of $Pr(F)$. This is the foundational work in this direction, and subsequently many researchers gave deep dive in this direction. 
\par Later Chen et al.  \cite{chen2010continuous} studied the approximate subgraph search problem in uncertain graph stream. Here, given an uncertain graph stream $<GS_{1}, GS_{2}, \ldots, GS_{k_{1}}>$, a set of query graphs $<Q_{1}, Q_{2}, \ldots, Q_{k_{2}}>$, and a probability threshold $\epsilon$ the goal here is to report all joinable pairs $<GS_{i,t} Q_{j}>$ in each time stamp $t$. This means that the subgraph $Q_{j}$ is contained in $GS_{i}$ with probability exceeding $\epsilon$, where $1 \leq i \leq k_{1}$, $1 \leq j \leq k_{2}$, and $t \geq 0$. They proposed two efficient pruning technique, namely `structural pruning' and `probability pruning' which makes there methodology efficient. Running time of their methodology is of $\mathcal{O}(\sqrt{n}m \log m)$.
 \par Later Yuan et al. \cite{yuan2011efficient2} studied the same problem and proposed the `filtering and verifying' strategy to speed up the search process. In the filtering phase, a probabilistic inverted index is maintained which can be used for probabilistic pruning. Next in the verification phase the remaining candidates have been verified by exact algorithm. This method is tested with both synthetic as well as real-world datasets. In a follow up work by yuan et al. \cite{yuan2014pattern}  \cite{yuan2016efficient}, they developed probabilistic match trees (PM Trees) based on match cuts and cut selection process. Considering this index structure, they developed effective pruning strategy to prune the unqualified matches. This makes the proposed methodology much more efficient, and hence the sizes of the graphs that have been used in this study is much larger than the previous works.

\par Chen et al. \cite{chen2018efficient} proposed the `enumeration-evaluation' framework for this problem, where first they enumerate all the candidate subgraphs and then for each subgraph compute its support and decide whether to output this as a result. They also showed that under the probabilistic semantic the computation of `Support' is \#P-Complete. Hence, their solutions are approximate in nature with accuracy guarantee. Experimental results show the practical usability of the algorithms.
\par Recently, Ma et al. \cite{ma2019linc} studied the problem of counting a given motif present in an uncertain graph. Given a motif $M$, there goal was to evaluate the occurrence statistics of $M$ such as probability mass function, mean, and variance. Based on their sample size analysis, they showed that if the  number of samples are more than $\frac{ln \frac{2}{\delta}}{2 \epsilon^{2}}$ then the absolute error is bounded by $\epsilon$ with probability $(1-\delta)$. The running time of their methodology is $\mathcal{O}(nmd^{|V_{M}|-2})$, where $|V_{M}|$ denotes the number of vertices in the motif and $d$ denotes the maximum degree of the graph. In the experimental study of this work the size of the datasets are much larger than that has been used in previous studies.
\paragraph{\textbf{Subgraph Similarity Search}} Yuan et al.  \cite{yuan2012efficient} \cite{yuan2015graph}  studied the subgraph similarity search problem over uncertain graph databases. They showed that the problem is $\#$P-Complete. They used their previously developed `filter and verify' technique to gear up the search process. In the filtering phase they develop lower and upper bound of subgraph similarity probability based on probabilistic matrix index. For the verification phase they developed efficient sampling approach to validate the remaining candidates. Experimental results show that their methods are scalable. 
\par Gu et al. \cite{gu2016subgraph} studied the problem of similarity maximal all matches in an uncertain graph database. Given an uncertain graph $G$, a query graph $g$, distance threshold $d$, and probability threshold $\alpha$, a deterministic graph $\mathcal{G} \in L(G)$ is a similarity maximal match of $q$ in $G$ under the vertex mapping $\mathcal{F}$, if there exist no other deterministic graph $\mathcal{G}^{'} \in L(G)$ such that under the same vertex mapping $\mathcal{F}$, $Pr(Dist(\mathcal{G}^{'},g) \leq d) > Pr(Dist(\mathcal{G},g) \leq d)$. Here, $Dist(\mathcal{G}^{'},g)$ is the edit distance between $\mathcal{G}^{'}$ and $g$ \cite{chen2019efficient}. They proposed different speed up techniques such as partial graph evaluation, vertex pruning, probability upper bound\mbox{-}based filtering, and the incremental evaluation method. Experimental results show that they outperform baseline methods in orders of magnitude.

Table \ref{Tab:Subgraph_1} summarizes the literature for pattern matching and subgraph similarity search queries on uncertain graphs.

\begin{table} 
\centering
	\begin{center}
	\caption{Summery of the literature of different structural pattern mining on uncertain graphs}
	\label{Tab:Subgraph_1}
		\begin{tabular}{ | p{2.7 cm} | p{2.7 cm} | p{6.4 cm} |}
			\hline
			\textbf{Type of Subgraph} & \textbf{Author} &  \textbf{Major Findings} \\
			\hline
			\multirow{2}{*}{Any subgraph } & Zou et al. \cite{zou2009frequent} & 	
\begin{itemize}
\item first to study the subgraph pattern matching queries on uncertain graphs.
\item derived a sample bound which is $\frac{16.n. \log(\frac{2}{\delta})}{\epsilon^{2}. minsup}$ for subgraph similarity search
\end{itemize}

			 \\
			\cline{2-3}
			& Chen et al.  \cite{chen2010continuous} & 
\begin{itemize}
\item proposed two efficient pruning technique, namely `structural pruning' and `probability pruning' for solving the problem.
\end{itemize}

			\\
			\cline{2-3} 
			& Yuan et al. \cite{yuan2011efficient2} \cite{yuan2014pattern}  \cite{yuan2016efficient} & 
\begin{itemize}
\item proposed `filtering and verifying' strategy to speed up the search process
\item developed probabilistic match trees (PM Trees) based on match cuts and cut selection process
\item developed effective pruning strategy to prune the unqualified matches
\end{itemize}

			\\
			\cline{2-3} 
			& Chen et al. \cite{chen2018efficient} & 
\begin{itemize}
\item proposed the `enumeration-evaluation' framework for this problem
\item showed that under the probabilistic semantic the computation of `Support' is \#P-Complete
\end{itemize}			
			
			\\
			\cline{2-3} 
			& Ma et al. \cite{ma2019linc} & 
\begin{itemize}
\item showed that if the number of samples are more than $\frac{ln \frac{2}{\delta}}{2 \epsilon^{2}}$, the absolute error is bounded
\item proposed a methodology with running time $\mathcal{O}(nmd^{|V_{M}|-2})$
\end{itemize}			
			
			\\
			\hline
			\multirow{2}{*}{Subgraph Similarity} & Yuan et al.  \cite{yuan2012efficient} \cite{yuan2015graph} &  
\begin{itemize}
\item showed that the subgraph similarity search problem is $\#$P-Complete
\item developed lower and upper bound of subgraph similarity probability based on probabilistic matrix index
\end{itemize}

			\\
			\cline{2-3}
			& Gu et al. \cite{gu2016subgraph} & 	
\begin{itemize}
\item proposed different speed up techniques such as partial graph evaluation, vertex pruning, probability upper bound-based filtering, and the incremental evaluation method.
\end{itemize}			
			
				 \\
			\hline
		\end{tabular}
	\end{center}
\end{table}

\paragraph{\textbf{Clique}} Clique in a deterministic graph $\mathcal{G}$ is defined as the subset of the vertices where every pair of them are adjacent to each other. A clique $\mathcal{C}$ is said to be maximal if for any $v \in V(\mathcal{G}) \setminus C$, $\mathcal{C} \cup \{v\}$ is not a clique. A clique is said to be a $k$-clique if the number of vertices in the clique is $k$. Zou et al. \cite{zou2010finding} studied the Top-$k$ maximal clique finding problem in uncertain graphs. They proposed a `branch and bound' algorithm for this problem. Given the uncertain graph $G$, first they construct the deterministic graph $\hat{G}$ by removing the uncertainties from $G$. Now, they defined a clique search tree where each node represents a distinct clique, the root be the trivial clique $\phi$, and if $\mathcal{C}$ is a parent of a non root clique $\mathcal{C}^{'}$ then $\mathcal{C} \subset \mathcal{C}^{'}$ with $|\mathcal{C}^{'}|=|\mathcal{C}|+1$. Now, the problem reduces to a tree searching problem and their algorithm is divided into the following steps: pruning, computing maximal clique probability, expansion, and update. Experimental results show that the methods are scalable. 
\par Later, Mukherjee et al. \cite{mukherjee2016enumeration} introduced the notion of an $\alpha$-maximal clique in the context of uncertain graphs and  studied counting and enumeration problem of such cliques. They showed that this number of an $n$ vertex uncertain graph is exactly $\binom{n}{\frac{n}{2}}$ and proposed an enumeration algorithm of such cliques having running time of $\mathcal{O}(n. 2^{n})$. They also showed that the running time of their proposed algorithm is $\mathcal{O}(\sqrt{n})$ of an optimal algorithm. experimental evaluation showing the practical utility of the algorithm. Experimental evaluation shows the practical utility of the algorithm. 
\par Recently, Li et al. \cite{li2019improved} studied the $(k,\tau)$-Clique search problem in an uncertain graph. A clique $\mathcal{C}$ is said to be a $(k,\tau)$-Clique if size of $\mathcal{C}$ is $k$ with clique probability at least $\tau$, and also $\mathcal{C}$ is maximal. They proposed two core based prunnning technique and one cut based prunning technique which makes the search process more effective. Running time of their algorithm is of $\mathcal{O}(2^{n^{'}}(m^{'} + n^{'}))$. Here, $m^{'}$ and $n^{'}$ denotes the number of edges and vertices of the largest connected component in $(Top_{k},\tau)$-core which is much smaller than the original uncertain graph. They defined the notion of $(Top_{k},\tau)$-core as the  Rashid et al. \cite{rashid2019top} 
 
\paragraph{\textbf{Truss}} In a deterministic graph $\mathcal{G}$, its one subgraph $H$ is called a $k$-truss if every edge of this subgraph is incident with at least $(k-2)$ tringles. Huang et al. \cite{huang2016truss} studied the truss decomposition problem in the context of uncertain graphs. They defined the notion of $(k, \gamma)$-truss as a maximal connected subgraph $\mathcal{H}$, in which for every edge, the probability that it is contained in at least $(k-2)$ triangles is at least $\gamma$. They proposed a dynamic programming algorithm for decomposing an uncertain graph into maximal such $(k, \gamma)$-truss. They also defined the global $(k, \gamma)$-truss in which along with the previous conditions, it has to satisfy that the probability that $\mathcal{H}$ contains a $k$-truss is at least $\gamma$.  
\par Subsequently, Zou and Zhu \cite{zou2017truss} introduced the $\theta$-truss decomposition problem ($\theta \in [0,1]$), where given an uncertain graph $G$, the goal is to decompose it into the following set $\{ T_{i, \theta}: 1 \leq i \leq k\}$. Here, $k$ is the largest value for which $T_{k,\theta}$ is a probabilistic truss. They showed that like deterministic graphs truss decomposition in uncertain graphs 
also follow `uniqueness' and `hierarchy' property. They proposed an $\mathcal{O}(m^{1.5}. \mathcal{Q})$ time algorithm to solve this problem. Here, $\mathcal{Q}$ is the maximum number of common neighbor of the end vertices of any edge. In the worst case, it may be of $\mathcal{O}(n)$. They also extend it to an external memory algorithm for this problem for larger size uncertain graphs. 
\paragraph{\textbf{Core}} In a deterministic graph $\mathcal{G}$, a subgraph $\mathcal{J}$ is said to be a $k$-core of $\mathcal{G}^{'}$ if $\forall v \in V(\mathcal{J})$, $deg(u)$ in $\mathcal{J}$ is more than or equal to $k$ and also $\mathcal{J}$ is maximal. Bonchi et al. \cite{bonchi2014core} introduced the notion of $(k, \eta)$-core, which is basically the maximal subgraph $\mathcal{J}$, such that every vertex $v \in V(\mathcal{J})$ has degree more than or equal to $k$ with probability at least $\eta$. First, they showed that given an $\eta$, the $(k,\eta)$-core decomposition of an uncertain graph is unique. They proposed a dynamic programming algorithm for this problem having run time of $\mathcal{O}(m \Delta)$, where $\Delta= \underset{v \in V(G)}{max} (\underset{k}{max} (Pr[deg(v) \geq k] \geq \eta))$. They applied this core decomposition approach to solve practical problems such as `influence maximization', `task driven team formation'. 
\par Later, Peng et al. \cite{peng2018efficient} introduced the concept of $(k,\theta)$-core of an uncertain graph which consists of set of nodes that is a $k$-core with probability  at least $\theta$. They showed that the problem of computing $(k,\theta)$-core is NP-Hard. If $\frac{1}{2 \epsilon^{2}}. ln(\frac{2n}{\delta})$ many graphs are sampled from the possible world then the estimation error of the $k$-core probability can be bounded by $\epsilon$ with probability at least $(1-\delta)$. To further reduce the computational burden in individual deterministic graphs, they proposed a $k$-core membership check algorithm that makes their methodology much faster.
 
 \paragraph{\textbf{Reliable Subgraph}} Jin et al. \cite{jin2011discovering} first defined the reliable subgraph search problem in the context of uncertain graphs, which basically asks given an $\alpha \in [0,1]$ determine all induced subgraphs of the input uncertain graph such that the probability that the remains connected is greater than or equal to $\alpha$. They showed if $\frac{2}{\epsilon^{2}} ln (\frac{2}{\delta})$ many number of deterministic graphs can be sampled from the possible world then the estimation error of the reliability of the induced subgraph by any subset of vertices $V_{s}$ can be bounded by $\epsilon$ with probability at least $(1-\delta)$. From this sampling scheme they defined a new problem called the `Frequent Cohesive Set problem'. A subset of the vertices of a graph is said to be cohesive if the subgraph induced by the vertex subset is connected. They developed a peeling algorithm which works from super sets to subsets during the pattern discovery. 
 
 \paragraph{\textbf{Nucleus}} Nucleus decomposition of a graph is generalization of core and truss decomposition \cite{sariyuce2015finding}. Recently, Esfahani et al.  \cite{esfahani2020nucleus} studied the problem of nucleus decomposition of an uncertain graphs. They have introduced the concept of `local' ($l$), `global'($gl$), and `weakly global' ($wg$) nucleus. Given an uncertain graph $G$, a triangle $\Delta$, and $\mu \in \{l,g,wg\}$, the probability that the random variable $X_{G, \Delta,\mu}$ takes the value $k$ with tail probability 
 \begin{equation}
 Pr[X_{G, \Delta,\mu} \geq k]= \underset{\mathcal{G} \in L(G)}{\sum} Pr_{\mathcal{G} \sqsubseteq G} . I_{\mathcal{G}, \Delta,\mu}
 \end{equation}
 Now, $I_{\mathcal{G}, \Delta,l}=1$ if $\Delta$ is in $\mathcal{G}$ and the support of $\Delta$ in $\mathcal{G}$ is at least $k$. $I_{\mathcal{G}, \Delta,gl}=1$ if $\Delta$ contained in $\mathcal{G}$, and $\mathcal{G}$ is a deterministic $k$ nucleus. They showed that the local version of the problem of nucleus decomposition is polynomial time solvable whereas the global version is NP-Hard. They proposed a dynamic programming approach for solving the local version. From the results they concluded that compared to probabilistic core and truss decompositions, nucleus decomposition significantly outperforms in terms of density and clustering metrics.   
 
 \par Table \ref{Tab:Subgraph_2} summarizes the literature related to different structural patterns in uncertain graph.

 \begin{table} 
	\begin{center}
	\caption{Summery of the literature of different structural pattern mining on uncertain graphs}
	\label{Tab:Subgraph_2}
		\begin{tabular}{ | p{1.5 cm} | p{2.7 cm} | p{6.4 cm} |}
			\hline
			\textbf{Type of Subgraph} & \textbf{Author} &  \textbf{Major Findings} \\
			\hline
			\multirow{2}{*}{Clique} & Zou et al. \cite{zou2010finding} & \begin{itemize}
			\item introduced the Top-$k$ maximal clique finding problem
			\item proposed a branch and bound algorithm for this problem
\end{itemize}			 \\
			\cline{2-3}
			& Mukherjee et al. \cite{mukherjee2016enumeration} & 
			\begin{itemize}
			\item introduced the notion of $\alpha$-maximal clique
			\item showed that the number of such cliques are $\binom{n}{\frac{n}{2}}$
			\item proposed an algorithm with running time $\mathcal{O}(n 2^{n})$ 
			\item showed that their algorithm is $\mathcal{O}(\sqrt{n})$ of optimal algorithm
			\end{itemize}\\
			\cline{2-3}
			& Li et al. \cite{li2019improved} & 
			\begin{itemize}
			\item introduced the notion of $(k, \tau)$-clique
			\item proposed core and cut\mbox{-}based pruning technique
			\item proposed an algorithm with running time of $\mathcal{O}(2^{n^{'}}(m^{'} + n^{'}))$
			\end{itemize}\\
			\hline
			\multirow{10}{*}{Truss} & Huang et al. \cite{huang2016truss} & \begin{itemize}
			\item introduced the truss decomposition problem in uncertain graphs
			\item proposed a dynamic programming algorithm
\end{itemize}			 \\
			\cline{2-3}
			& Zou and Zhu \cite{zou2017truss} & \begin{itemize}
			\item introduced the $\theta$-truss decomposition problem
			\item showed that the truss decomposition follow `uniqueness' and `hierarchy' property
			\item proposed an $\mathcal{O}(m^{1.5}. \mathcal{Q})$ time algorithm
\end{itemize}			 \\
			\hline
			\multirow{10}{*}{Core} & Bonchi et al. \cite{bonchi2014core} & \begin{itemize}
			\item introduced the core decomposition problem in uncertain graphs
			\item proposed an algorithm with running time $\mathcal{O}(m \Delta)$
			\item applied there approach to solve `influence maximization' and `task driven team formation'
\end{itemize}			 \\
			\cline{2-3}
			& Peng et al. \cite{peng2018efficient} & \begin{itemize}
			\item introduced the concept of $(k, \theta)$-core
of an uncertain graph
			\item gave a sample bound to compute the $k$-core probability with bounded error
			\item proposed a $k$-core membership check algorithm for speeding up the procedure
\end{itemize}			 \\
			\hline
			\multirow{10}{*}{Reliable } & Jin et al. \cite{jin2011discovering} & \begin{itemize}
			\item first defined the problem of reliable subgraph search on uncertain graphs
			\item gave a sample bound to estimate the subgraph reliability with bounded error
\end{itemize}			 \\
			\hline
			\multirow{10}{*}{Nucleus } & Jin et al. \cite{jin2011discovering} & \begin{itemize}
			\item first defined the problem of reliable subgraph search on uncertain graphs
			\item gave a sample bound to estimate the subgraph reliability with bounded error
\end{itemize}			 \\
\hline
		\end{tabular}
	\end{center}
\end{table}
\subsection{Graph Algorithmic Problem}
As mentioned in Section \ref{Sec:Problems}, among plethora of graph algorithmic problems, only a very few has been studied on uncertain graphs. The first one is the construction of `Minimum Spanning Tree (MST).' Zhang et al. \cite{zhang2016minimum} introduced and studied the problem of finding most reliable minimum spanning tree. The trivial solution for this problem is to enumerate all the possible worlds, compute the MST in each one of them and then choose one with the highest probability value. However, it requires exponential time. To get rid of this problem, Zhang et al. \cite{zhang2016minimum} proposed an incremental greedy approach for this problem. Starting with an empty MST, until all the vertices of the graph has been included in the MST this algorithm iteratively choose an edge that connects the already build MST with maximum probability. To implement this, they have used the `max heap' data structure. Their analysis shows that this algorithm is able to give a spanning tree having reliability $(1-(\frac{1}{2})^{\frac{d}{2}})^{n-1}$ and expected approximation ratio $(\frac{1}{2})^{dn}$, where $d$ is the maximum degree of the graph. Computational time requirement of this algorithm is $\mathcal{O}(d^{2}.n^{2})$. To the best of our knowledge, there is no other study that on finding MST on uncertain graphs.
\par Another important problem which has been studied in the context of uncertain graphs is the `link prediction problem'. Ahmed et al.  \cite{ahmed2016efficient} studied this problem in uncertain social networks. They reduced the link prediction problem in uncertain network to the problem of random walk in a deterministic graph. They considered the `Simrank' as the similarity metric between two nodes. Their method consisting of the following two steps: 
\begin{itemize}
\item First step is the computation of the transformation matrix. For the given probabilistic graph $G(V,E,P)$ and a deterministic graph $\mathcal{G}(V, \mathcal{E})$, the associated transformation matrix can be defined as follows:
\begin{equation}
\overline{W}(u,u)= \underset{(uq) \in E(\mathcal{G})}{\prod} (1-P(uq))
\end{equation}
\begin{equation}
\overline{W}(u,v)=\underset{G^{'} \in \Omega(uv)}{\sum} \frac{A(u,v)}{\underset{(uq) \in E(G^{'})}{\sum} A(u,q)}   Pr_{G^{'} \sqsubseteq G}
\end{equation}
Where $A$ is the adjacency matrix of the uncertain graph $\mathcal{G}$, and $\Omega(uv)$ is the subset of the possible worlds in which the edge $(u,v)$ is present.
\item Second step is the computation of the Simrank matrix. Based on the computed value of $\overline{W}$, this matrix is computed using the following equation:
\begin{equation}
S=c.\overline{W}^{T}.S.\overline{W}+(1-c).I
\end{equation}
where, c is a constant having value in between $0$ and $1$, and $I$ is the identity matrix. 
\end{itemize}  
Based on this computed Simrank value, for a given $k \in \mathbb{Z}^{+}$, vertex pairs (which are not currently linked currently) corresponding top-$k$ values are returned as the result.
\par Another algorithmic problem studied in the context of uncertain graph is the `information flow maximization' problem. Frey et al.  \cite{frey2017efficient, DBLP:conf/icde/FreyZER18} showed that this problem is $NP$-Hard by a reduction from the \emph{0-1 Knapsack Problem} \cite{moradi2021efficient}. To highlight, they study this problem on vertex weighted graphs, where vertex weight signifies the weight on the information in the corresponding vertex. For a vertex $Q \in V(G)$,  $flow(Q,G)$ denotes the random variable signifying the total vertex weights of all the nodes $V(G)$ that are reachable from the node $Q$. As mentioned previously, rechability problem in uncertain graphs is $\#P$-Hard, hence the complexity of the information flow estimation also falls in the same category. To tackle this problem, they used traditional Monte Carlo Sampling to generate a subset of possible worlds. It has been shown that the average value over these samples is an unbiased estimator of the random variable $flow(Q,G)$. Next, they have used the concept mono-connected nodes (two nodes are mono-connected if there exist a single path connecting them and also for them the reliability can be computed efficiently.), and bi-connected nodes (two nodes are mono-connected if there exist two or more path connecting them). Next, they proposed the \emph{Flow Tree} data structure which is an adoption of \emph{Block Cut Tree} \cite{hopcroft1973algorithm, westbrook1992maintaining}. As, finding the optimal solution of this problem is $NP$-Hard, hence they proposed an incremental greedy algorithm that generates sub-optimal solution for this problem. 
\subsection{Miscellaneous Problem}
Among a few other problems, `graph sparcification' has been studied in the context of uncertain graphs. As mentioned previously, for a given sparcification ratio $\alpha \in (0,1)$ and an uncertain graph $G(V, E, P)$ , this problem is all about find a subgraph $G^{'}(V, E^{'}, P^{'})$ such that $E^{'} \subset E$ and $|E^{'}| =\alpha. |E|$. For this problem Parchas et al. \cite{parchas2018uncertain} proposed a cut\mbox{-}based sparcification technique for uncertain graphs. First, they generalized the notion of cut size for a given set of vertices $S$ in the context of uncertain graph. Their proposed framework begins by initializing a connected unweighted backbone graph $G_{b}$ and after that the two techniques are applied to $G_{b}$ to generate the spercified graph. First,  they proposed an approach to initialize the backbone graph $G_{b}(V, E_{b})$. As described in Section \ref{SubSubSec:Sparcification}, the problem is to minimize the discrepancy, hence they formulated the following optimization problem:
\begin{center}
$\underset{p^{'}}{min} |d-Y_{b}. p^{'}|$\\
such that, $p^{'} \in (0,1]^{|E_{b}|}$
\end{center}
where $d$ is a vector of size $|V(G)|$ containing the expected degree values of the nodes of $G$ and $Y_{b}$ is the incidence matrix of size $|V(G)| \times E_{b}$. Next, they showed that for the incidence matrix $A_{b}$ of $G_{b}$, there exist a probability assignment $p^{*}$ that minimizes the discrepancy for which the expected degree of any node is less than its original degree. Using this result they showed that the optimal probability distribution $p^{*}$ for the degree discrepancy $\delta_{A}$ is the solution of the following linear programming problem (LPP):\\
\begin{center}
$\underset{p^{'}}{max} \ |p^{'}|$ \\
such that, $A_{b}.p^{'} \leq d$ \\
and $p^{'} \in (0,1]^{|E_{b}|}$.
\end{center}
Now, this LPP can be solved using the solver packages like, \emph{Simplex} \cite{ficken2015simplex}. They also proposed an efficient methodology `Gradient Decent Backbone' which closely approximates the optimal probability assignment. The running time of this methodology is of $\mathcal{O}(M_{steps}. \alpha. |E(G)|)$, where $M_{steps}$ denotes the number of steps required to converge the gradient decent method. This method only updates the edge probabilities without inserting or removing edges. They proposed another methodology called \emph{Expectation Maximization Degree} and this modifies both the backbone graph as well as the edge probabilities. This procedure is inspired by the \emph{expectation-maximization}, which is an iterative optimization framework and also uses the Gradient Decent Backbone as a subroutine. Running time of this methodology is that $\mathcal{O}(E_{steps}. \alpha. |E(G)|. (\log n + M_{steps}))$, where $E_{steps}$ denotes the number of iterations required to converge this process.
\par Another problem that has been studied in the context of uncertain graph is the `node classification' problem. Dallachiesa et al.  \cite{dallachiesa2014node} proposed an iterative probabilistic learning approach to solve this. For any unlabelled node $r$ in $G$ having neighbors $i_{1}$, $i_{2}$, $\ldots, i_{s}$ with labels $t_{1}$, $t_{2}$, $\ldots, t_{s}$, its probability is estimated by the Bayes' rule over the adjacent nodes' neighbors. This is formalized by the following equation:
\begin{equation}
P(L(r)=p|L(i_1)=t_1, \ldots,L(i_s)=t_s) \propto P(L(r)=p) \underset{x}{\prod} P(L(i_x)=t_x|L(r)=p)
\end{equation}
However, if significant number of edges have low probability, then this method will have an impact on the classification process. To get rid of this problem, they proposed the iterative edge augmentation process. For this purpose, they split the already labeled nodes ($T_{0}$) into two parts `training' ($T_{train}$) set and a `hold out' set ($T_{hold}$). The ratio $T_{train}$ to $T_{0}$ is basically a user defined parameter. Initially, starting with a small fraction of high probability edges, this edge set is expanded with incorporating the outside edges in an iterative manner. the ratio of active edges are denoted by $\theta$. The value of $\theta$ for which maximum accuracy is obtained is denoted by $\theta^{*}$. This can be obtained by evaluating the accuracy on a number of sampled graphs.
\par Han et al. \cite{han2015uncertain} studied the uncertain graph classification problem using the \emph{extreme learning machine}, which was originally proposed by Miche et al. \cite{miche2009op}. Their method is broadly divided into three steps. Firstly, we put forward a framework for classifying uncertain graph objects. Secondly, we extend the traditional algorithm used in the process of extracting frequent subgraphs to handle uncertain graph data. Thirdly, based on Extreme Learning Machine (ELM) with fast learning speed, a classifier is constructed.
\section{Current Trends and Exsisting Research Gaps} \label{Sec:Gaps}
After analyzing the reported literature, here we list out the current research trends:
\begin{itemize}
\item \textbf{Bounds on Required Samples:} As mentioned previously, for most of the problems on uncertain graph how we can sample few of them to answer the problem within bounded error with high probability. Recently,  sample bounds are analyzed for different problems studied on uncertain graphs. Sadeh et al. \cite{sadeh2019sample} studied this for the influence maximization problem. However, the same study can be done for other problems as well. 
\item \textbf{Development of Efficient Indexing:} Another way of reducing the computational time of processing any uncertain graph is to index it properly using the proposed indexing scheme. To the best of our knowledge, the only indexing scheme is available for uncertain graphs is \emph{probabilistic tree} \cite{maniu2017indexing}. Much more indexing scheme can be developed which may lead to faster execution of the query problems on uncertain graphs. 
\item \textbf{Finding Different structural Patterns:} Recently several structural patterns finding algorithms are developed for analyzing the topology. As an example, Sun et al. \cite{sun2021efficient} proposed a methodology for indexing the trusses of an uncertain graph. Such methods are not available in the literature for other structural patterns.
\item \textbf{Uncertain Graph Sparsification:} A Plethora of literature is available for sparsification of a deterministic graph \cite{peleg1989graph, althofer1993sparse, fung2019general}. Recently, a cut-based spercification technique has been developed for uncertain graphs. Certainly, there is a gap in this area as the other principles of graph sparsification can be used for designing the same for the uncertain graph.
\item \textbf{Graph Algorithmic Problems:} As mentioned previously, among plethora of graph algorithmic problems, in last five years very few of them has been studied such as spanning tree, link prediction, information flow maximization. There is a research gap as the remaining algorithmic graph theory problems can be studied under the uncertain graph framework.  
\end{itemize} 
Next, we proceed to describe the future research directions in this domain.
\section{Future Research Directions} \label{Sec:ReDir}
Based on the detailed analysis of the existing literature, in this section, we describe a number of future research directions.
\begin{itemize}
\item \textbf{Definition of Uncertainty}: In the existing literature, the notion of uncertainty is defined as the probabilistic existence of the edges of the graph. However, it can be observed that this not the only notion of uncertainty. The same thing applies to the vertices as well. Hence all the problems mentioned in Section \ref{Sec:Problems}, that has been under edge uncertainty model, can also be studied under node uncertainty model as well. Very recently, Fukunaga et al.   \cite{fukunaga2019adaptive} studied a well studied graph theoretic problem \textsc{Connected Dominating Set Problem} under the node uncertainty setting.
\item \textbf{Modeling the Problem}: Before applying any principles and approaches of uncertain graph mining, it is important to represent the input network as an uncertain graph, and for that it is very much vital to decide which uncertain model is going to fit.
\item \textbf{Studying Existing Graph Theoretic Problems in Uncertain Graph Framework}: There are tons of graph theoretic problems which has been studied by theoretical computer scientist. However, as mentioned in Section \ref{Sec:Problems}, very few of them have been studied in uncertain graph framework. So, naturally it appears that there are many problems which has not been studied yet in uncertain graph framework. it will be interesting to take an existing problem and study in uncertain graph framework. Also, different model of uncertainty may give altogether different flavor to the problem. 
\item \textbf{Efficient Sampling Scheme}: There are many problems in uncertain graph mining domain, such as rechability (i.e., reliability computation), subgraph enumeration and many more, where sampling of the deterministic graphs from the possible worlds play a crucial role. It will be very important contribution to come up with efficient sampling technique, that can beat the existing techniques in atleast any one of the following two metrics: either (i) it should take less number of samples compared to the existing methods, or (ii) it should be able to answer to the queries with higher probability.
\item \textbf{Information Flow Maximization Problem}: In  Frey et. al.'s \cite{frey2017efficient} study of this problem, they have only consider the Monte-Carlo Sampling. However, as mentioned previously, there are many other and more efficient sampling techniques are there. Those sampling techniques can be used solve the problem even more efficiently. 
\item \textbf{Incorporation of Temporal Nature}: In almost entire existing literature, it has been considered that the assigned probability values to the edges are fixed over the time. However, in reality the case is not the same. As mentioned previously, the influence graph of a social network often represented as an uncertain graph.  The links of a social network are time varying, and also if the probability values marked on the edges are considered as the diffusion probability that can also change with time \cite{laurent2015calls}. 
\item \textbf{Querying of Combinatotial Structures:} As mentioned in Section \ref{Sec:Querying_Problems}, finding and counting motifs in networks is important problem. There are different kinds of motifs that are available in network science literature, such as \emph{Clan}, \emph{Core}, \emph{Club} and their different generalizations. However, it is surprising to observe that there are very few structures which have been studied in uncertain graph framework. So, it will be an interesting future work to develop efficient algorithms for finding remaining structural patterns in an uncertain graphs.
\item \textbf{Solution Methodologies for Modern Hardware}: Recent times have witnessed a significant development in almost every aspects computer architecture, starting from processor and accelerator design to memory system or subsystem organization \cite{deng2017toward, kloda2019deterministic}. Also, several new computing paradigms have been introduced, such as \emph{map reduce} and many more. For deterministic graph mining there exist some literature, which developed algorithms for modern computing paradigms \cite{xiang2013scalable, chen2016parallelizing}. Similarly, for uncertain graph mining also algorithms can be developed in the context of modern computing paradigms. 
\item \textbf{Scalability of the Methodologies}: Day by day the size of the real\mbox{-}world networks are increasing. As an example, it can be observed that for the social influence maximization problem (a problem in the domain of social network analysis, where the underlying social network is represented as an uncertain graph) the experimentation of the very first paper by Kempe et al. \cite{kempe2003maximizing} uses dataset of size $10748$ nodes and $53000$ edges. However, a very recent paper on the same problem by Zeng et al. \cite{zeng2020rcelf} uses dataset of size $41.7 \times 10^{6}$ nodes and $1.5 \times 10^{9}$ edges. So, to deal with this kind of gigantic networks, it is essential that the solution methodologies should have high scalability. Hence, developing highly scalable methods for large scale uncertain graphs will remain major thrust in this area.

\item \textbf{Privacy Preserving Uncertain Graph Mining}: As mentioned previously, the privacy preserving aspects has been deeply inducted with data mining in recent times. However, in the context of uncertain graphs this study is extremely limited \cite{tian2018privacy, xiao2018sharing}. So, more research work can be carried out in this direction.
\item \textbf{System Oriented Research}: Since last one decade or so, a significant effort has been made to develop graph processing systems \cite{zhu2016gemini, malewicz2010pregel}. Please look into \cite{mccune2015thinking, shi2018graph, besta2019graph}. However, none of the studies consider the uncertain nature of the graph. So, it will be an interesting future work to develop uncertain graph processing and mining system, which includes efficient data structures for storing, efficient processing algorithm developed based on processor architecture, and also very effective data visualization, and interpretation component to represent the output.
\end{itemize}
\section{Concluding Remarks} \label{Sec:Conclusion}
In this paper, we have presented a concise and almost self contained survey on the uncertain graph mining and analysis with a major focus on three aspects, namely, different problems studied in this domain, challenges for solving those problems, and existing solution methodologies. After analyzing the existing literature, we have derived the current research trend. At the end, we have listed a number of future research directions. Hope that this survey will help the upcoming researchers and practitioners to have finer understanding and better exposure in this field.
%
%


%
%

\bibliographystyle{spbasic}      
\bibliography{Paper}   


\end{document}